\newcommand{\beq}{\begin{equation}}
\newcommand{\eeq}{\end{equation}}
\newcommand{\bea}{\begin{eqnarray}}
\newcommand{\eea}{\end{eqnarray}}

\newcommand{\gsim}{\lower.7ex\hbox{$\;\stackrel{\textstyle>}{\sim}\;$}}
\newcommand{\lsim}{\lower.7ex\hbox{$\;\stackrel{\textstyle<}{\sim}\;$}}

\newcommand{\mrm}{\mathrm}

\documentclass[aps,prev,twocolumn,preprintnumbers,floatfix,nofootinbib]{revtex4-1}
\usepackage{graphicx}
\usepackage{epstopdf}
\usepackage{mathrsfs}
\usepackage{amssymb}
\usepackage{verbatim}

\def\stacksymbols #1#2#3#4{\def\theguybelow{#2}
    \def\vp{\lower#3pt}
    \def\sp{\baselineskip0pt\lineskip#4pt}
    \mathrel{\mathpalette\intermediary#1}}

\def\intermediary#1#2{\vp\vbox{\sp
     \everycr={}\tabskip0pt
     \halign{$\mathsurround0pt#1\hfil##\hfil$\crcr#2\crcr
              \theguybelow\crcr}}}

\def\be{\begin{equation}}
\def\ee{\end{equation}}
\def\bea{\begin{eqnarray}}
\def\eea{\end{eqnarray}}

\def\sp{\;\;\;,\;\;\;}

\def\mrm{\mathrm}

\def\lsim{\raise0.3ex\hbox{$\;<$\kern-0.75em\raise-1.1ex\hbox{$\sim\;$}}}
\def\gsim{\raise0.3ex\hbox{$\;>$\kern-0.75em\raise-1.1ex\hbox{$\sim\;$}}}

\def\inbar{\,\vrule height1.5ex width.4pt depth0pt}

\def\IC{\relax\hbox{$\inbar\kern-.3em{\rm C}$}}
\def\IQ{\relax\hbox{$\inbar\kern-.3em{\rm Q}$}}
\def\IR{\relax{\rm I\kern-.18em R}}
 \font\cmss=cmss10 \font\cmsss=cmss10 at 7pt
\def\IZ{\relax\ifmmode\mathchoice
 {\hbox{\cmss Z\kern-.4em Z}}{\hbox{\cmss Z\kern-.4em Z}}
 {\lower.9pt\hbox{\cmsss Z\kern-.4em Z}}
 {\lower1.2pt\hbox{\cmsss Z\kern-.4em Z}}\else{\cmss Z\kern-.4em Z}\fi}

\def\comment#1{}

\def\u1x{U(1)_X}
\newcommand{\nc}{\newcommand}
\nc{\LL}{L}
\nc{\vv}{\tilde{v}}
\nc{\ccdot}{\!\cdot\!}
\nc{\gsm}{G_{SM}}
\nc{\vfive}{\mathbf{5}\oplus\mathbf{\overline{5}}}
\nc{\vten}{\mathbf{10}\oplus\mathbf{\overline{10}}}
\nc{\zhol}{Z^{\rm hol}}
\nc{\xfb}{\,{\rm fb}}

\begin{document}

%
%

\preprint{LPT--Orsay 10/46}

\title{The ZZ' kinetic mixing in the light of the recent direct and indirect dark matter searches}

\author{Yann Mambrini$^{a}$}
\email{Yann.Mambrini@th.u-psud.fr}

\vspace{0.2cm}
\affiliation{
${}^a$ Laboratoire de Physique Th\'eorique \\
Universit\'e Paris-Sud, F-91405 Orsay, France}

\begin{abstract}

Several constructions, of stringy origins or not, generate abelian gauge extensions of the Standard Model (SM). 
Even if the particles of the SM are not charged under this extra $U'(1)$, one cannot avoid
the presence of a kinetic mixing between $U'(1)$ and the hypercharge $U_Y(1)$.
In this work, we constraint drastically this kinetic mixing, taking into account the recent experimental data  from
accelerator physics, direct detection and indirect detection of dark matter. We show that the region respecting
WMAP and experimental constraints is now very narrowed along the pole line where $M_{Z_D}\simeq 2 m_{DM}$,
$Z_D$ being the gauge boson associated to the extra $U'(1)$.

\end{abstract}

\maketitle


\maketitle


\setcounter{equation}{0}



\section{Introduction}

Neutral gauge sectors with an additional dark $U_D(1)$ symmetry in addition 
to the Standard Model (SM) hypercharge $U(1)_Y$ and an associated $Z_D$ 
("D" standing for $Dark$)
are among the best motivated extensions of the SM, and give the possibility
that a dark matter candidate lies within this new gauge sector of the theory.
Extra gauge symmetries are predicted in most Grand Unified Theories (GUTs)
and appear systematically in string constructions. Larger groups than $SU(5)$
or $SO(10)$, like $E_6$ allows the SM gauge group to be embedded into them.
 Brane--world $U'(1)$s are special compared to GUT $U'(1)$s because there 
 is no reason for the SM particle to be charged under them:
for a review of the phenomenology of the extra $U'(1)$s generated in such scenarios see e.g.
 \cite{Langacker:2008yv}.
In such framework, the extra--gauge boson would act as a portal between the $dark$ world
(particles not charged under the SM gauge group) and the $visible$ sector through its gauge 
invariant kinetic mixing $\delta/2 F_Y^{\mu \nu} F^D_{\mu \nu}$ 
\cite{Holdom, Feldman:2007wj,Dienes:1996zr,Martin:1996kn,Rizzo:1998ut,
delAguila:1995rb,Dobrescu:2004wz,Cohen:2010kn,Kang:2010mh}.
One of the first model of dark matter from the hidden sector with a massive additional
$U'(1)$ through both mass and kinetic mixings, the so called dark force
can be found in  \cite{Feldman:2006wd}.
The Dark Matter (DM) candidate $\psi_0$ would be the lightest (and thus stable) particle of
this secluded sector.
Such a mixing has been justified in
recent string constructions \cite{Cicoli:2011yh,Kumar:2007zza,
Javier,Cassel:2009pu}, 
but has also been studied with a model independent
approach \cite{Feldman:2007wj,Chun:2010ve,Pospelov:2008zw,Mambrini:2010yp}.

On the other hand, recently the CRESST-II and CoGENT collaborations have reported the observation of 
low--energy events in excess of known backgrounds \cite{CRESST, COGENT}. This
has encouraged the hypothesis that these signals  -in addition to the long--standing
DAMA \cite{DAMA} annual modulation signal -- might arise from the scattering of a light
($\sim 10$ GeV) dark matter particle. It was recently shown that models  with extra $U'(1)$
can easily accommodate with such signals
 \cite{Mambrini:2010dq,Hooper:2010uy,Cheung,Fitzpatrick:2010em}.
 Similar models with singlet extension reached the same conclusion because they are very similar
 in the construction than the models with extra gauge boson, except that the role of the
 portal is played by the Higgs boson \cite{Tytgat, Tytgatbis}.
 Other $very$ important astrophysical  consequences  from the tri--bosonic coupling $Z_DYY$
generated by loop induced triangle diagrams  are
studied in \cite{Mambrini:2009ad,Dudasline,Higgsspace} and in \cite{Vertongen:2011mu} in a more
model independent way. Consequences of such anomaly-generated vertex are also well
analyzed in \cite{Kumar:2006gm,Antoniadis:2009ze}. It is worth noticing that one can easily build a model
where the $Z_D$ becomes the DM candidate \cite{Hambye:2008bq}:
a clear summary of all these extensions can be found in \cite{Hambye:2010zb}.

Our objective in this work is to restrict the parameter space of an extra $U_D(1)$, taking into consideration
the main detection modes, $i.e.$ cosmological data, precision measurements, direct and indirect detection
searches of dark matter. The paper is organized as followed: after a brief reminder of the model and its
phenomenological characteristics, we analyze the experimental exclusion limits (in a conservative way)
in different regimes ((very)light and (very)heavy DM), before presenting a general analysis and perspective
for a XENON--like 1 ton detector. We then conclude.

\section{The model}


The matter content of any $dark$ $U(1)_D$ extension of the SM can be decomposed
into three families of particles:

\begin{itemize}
\item{The $Visible$ $sector$ is made of particles which are charged under the SM
gauge group $SU(3)\times SU(2)\times U_Y(1)$ but not charged under $U_D(1)$
(hence the $dark$ denomination for this gauge group)}
\item{the $Dark$ $sector$ is composed by the particles charged under
$U_D(1)$ but neutral with respect of the SM gauge symmetries. The dark matter
($\psi_0$) candidate is the lightest particle of the $dark$ $sector$}
\item{The $Hybrid$ $sector$ contains states with SM $and$ $U_D(1)$ quantum numbers. These states are fundamental because they act as a portal between
the two previous sector through the kinetic mixing they induce at loop
order.} 
\end{itemize}

\noindent
From these considerations, it is easy to build the effective Lagrangian
generated at one loop :

\begin{eqnarray}
{\cal L}&=&{\cal L}_{\mrm{SM}}
-\frac{1}{4} \tilde B_{\mu \nu} \tilde B^{\mu \nu}
-\frac{1}{4} \tilde X_{\mu \nu} \tilde X^{\mu \nu}
-\frac{\delta}{2} \tilde B_{\mu \nu} \tilde X^{\mu \nu}
\nonumber
\\
&+&i\sum_i \bar \psi_i \gamma^\mu D_\mu \psi_i
+i\sum_j \bar \Psi_j \gamma^\mu D_\mu \Psi_j
\label{Kinetic}
\end{eqnarray}

\noindent
$\tilde B_{\mu}$ being the gauge field for the hypercharge, 
$\tilde X_{\mu}$ the gauge field of $U_D(1)$ and
$\psi_i$ the particles from the hidden sector, $\Psi_j$ the particles
 from the hybrid sector, 
$D_{\mu}  =\partial_\mu -i (q_Y \tilde g_Y \tilde B_{\mu} + q_D \tilde g_D
 \tilde X_{\mu} + g T^a W^a_{\mu})$, $T^a$ being the $SU(2)$ generators, and 

\beq
\delta= \frac{\tilde g_Y \tilde g_D}{16 \pi^2}\sum_j q_Y^j q_D^j 
\log \left( \frac{m_j^2}{M_j^2} \right)
\eeq

\noindent
with $m_j$ and $M_j$ being hybrid mass states \cite{Baumgart:2009tn} .

Notice that the sum is on all the hybrid states, as they are the only ones which can contribute to the
 $\tilde B_{\mu} \tilde X_{\mu}$ propagator.
After diagonalization of the current eigenstates that makes the gauge kinetic
terms of Eq.(\ref{Kinetic}) diagonal and canonical, 
we can write after the $SU(2)_L\times U(1)_Y$ breaking\footnote{Our notation
for the gauge fields are 
($\tilde B^\mu,\tilde X^\mu$) before the diagonalization, 
($B^\mu, X^\mu$) after diagonalization and 
($Z^\mu,Z_D^\mu$) after the electroweak breaking.} :

\begin{eqnarray}
A_{\mu} &=& \sin \theta_W W_{\mu}^3 + \cos \theta_W B_{\mu}
\\
Z_{\mu} &=& \cos \phi ( \cos \theta_W W_{\mu}^3 - \sin \theta_W B_{\mu})
- \sin \phi  X_\mu
\nonumber
\\
(Z_D)_{\mu}&=&\sin \phi (\cos \theta_W W_\mu^3 - \sin \theta_W B_\mu)
+ \cos \phi  X_\mu
\nonumber
\end{eqnarray}

\noindent
with, at the first order in $\delta$:

\begin{eqnarray}
\cos \phi &=& \frac{\alpha}{\sqrt{\alpha^2 + 4 \delta^2 \sin^2 \theta_W}}
~~
\sin \phi = \frac{2 \delta \sin \theta_W}{\sqrt{\alpha^2 + 4 \delta^2 \sin^2 \theta_W}}
\nonumber
\\
\alpha &=& 1- M^2_{Z_D}/M^2_Z - \delta^2 \sin^2 \theta_W
\label{sphi}
\\
&\pm& \sqrt{(1-M^2_{Z_D}/M^2_Z -\delta^2 \sin^2 \theta_W)^2+ 4 \delta^2 \sin^2 \theta_W}
\nonumber
\end{eqnarray}

\noindent
and + (-) sign if $M_{Z_D}< (>)M_Z$.
The kinetic mixing parameter $\delta$ generates an effective coupling of 
SM states $\psi_{\mrm{SM}}$ to $Z_D$, and a coupling of $\psi_0$ to 
the SM $Z$ boson which
induces an interaction on nucleons.
Developing the covariant derivative on SM and $\psi_0$ fermions state,
we computed the effective $\psi_{\mrm{SM}}\psi_{\mrm{SM}}Z_D$  and
 $\psi_0\psi_0Z$ couplings at first order\footnote{  One can find
  a detailed analysis of the spectrum and couplings of the model
   in the appendix of \cite{Chun:2010ve}.} in $\delta$ and obtained
   
   \bea
   {\cal L}= q_D \tilde g_D (\cos \phi~ Z'_\mu \bar \psi_0 \gamma^{\mu} \psi_0 +
   \sin \phi~ Z_\mu \bar \psi_0 \gamma^{\mu} \psi_0 ).
   \eea

 \noindent
 We took $q_D \tilde g_D=3$ trough our analysis, keeping in mind that our results stay 
 completely general by a simple rescaling of the kinetic mixing\footnote{The author would like to thank the referee for having pointed out this detail of the analysis.} $\delta$.

\begin{figure}
    \begin{center}
   \includegraphics[width=3.in]{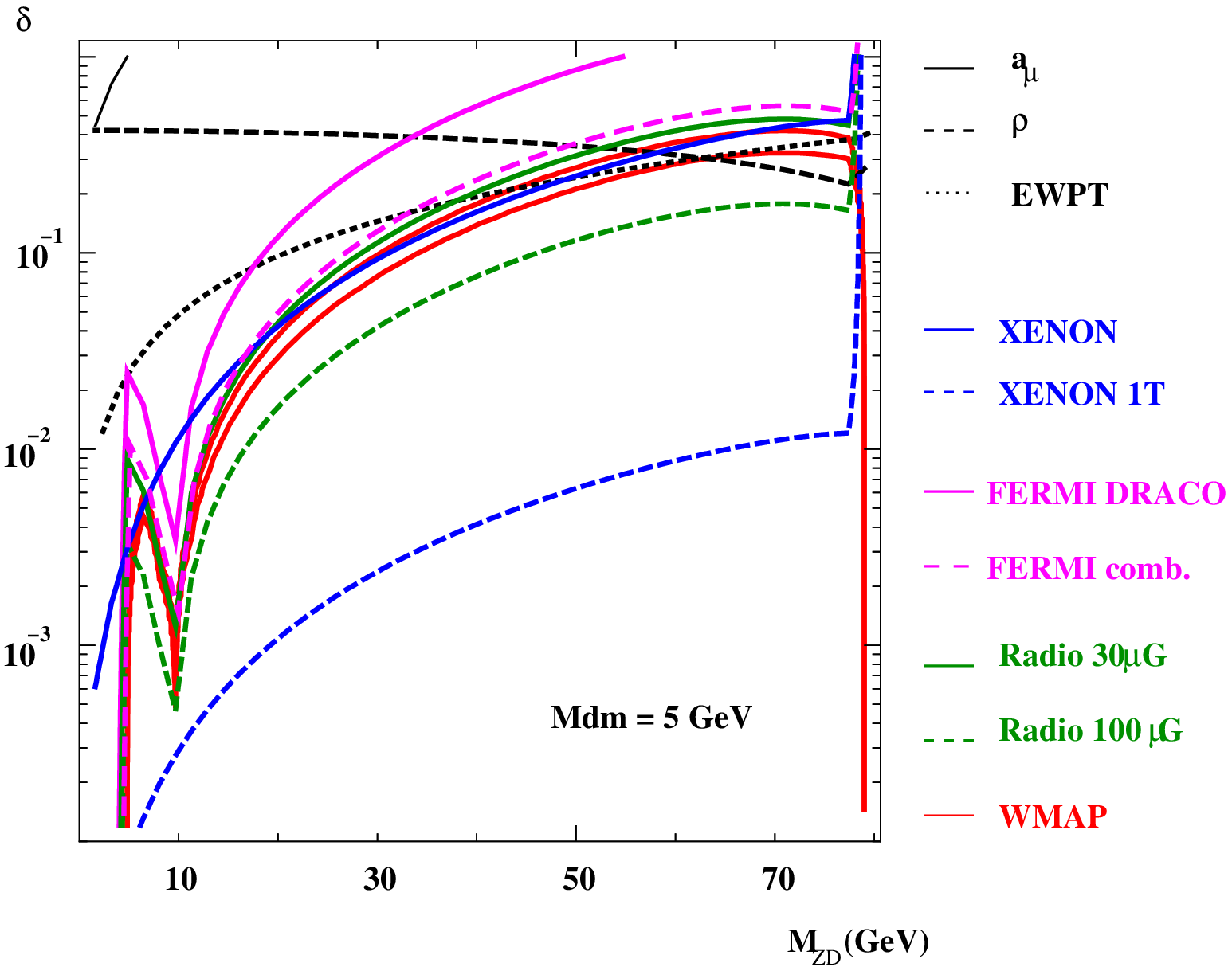}
   
    \includegraphics[width=3.in]{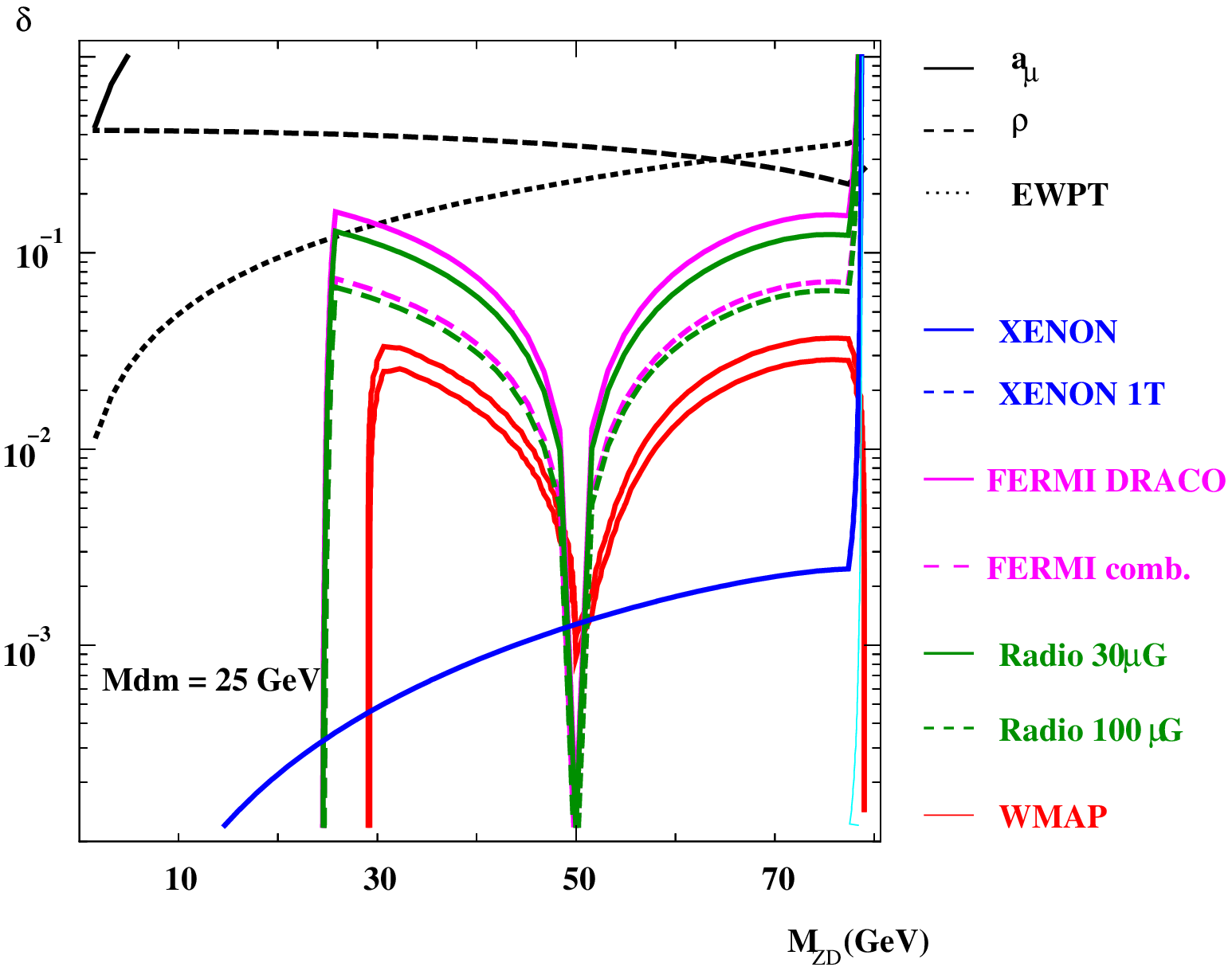}
    
    \includegraphics[width=3.in]{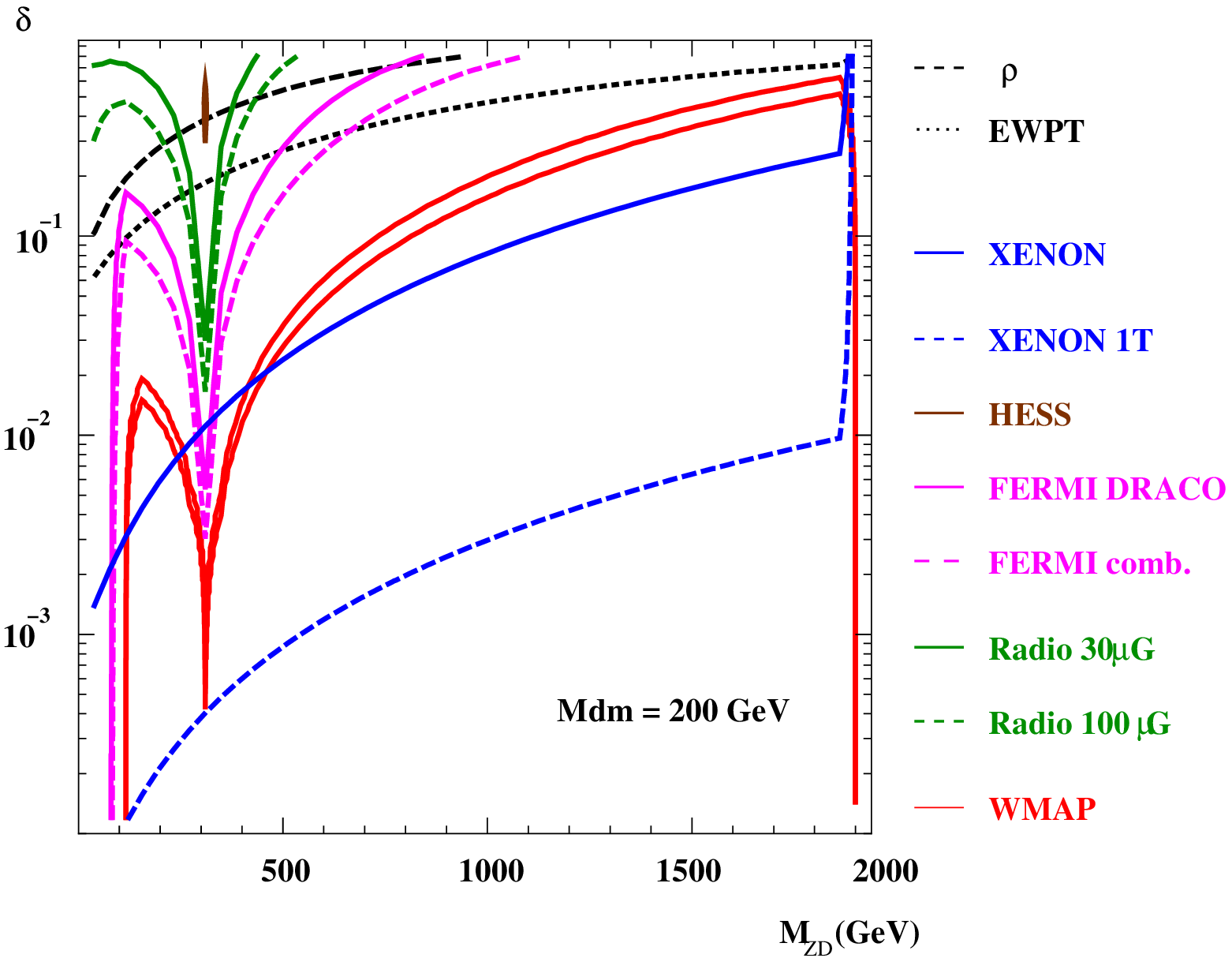}
    
        \includegraphics[width=3.in]{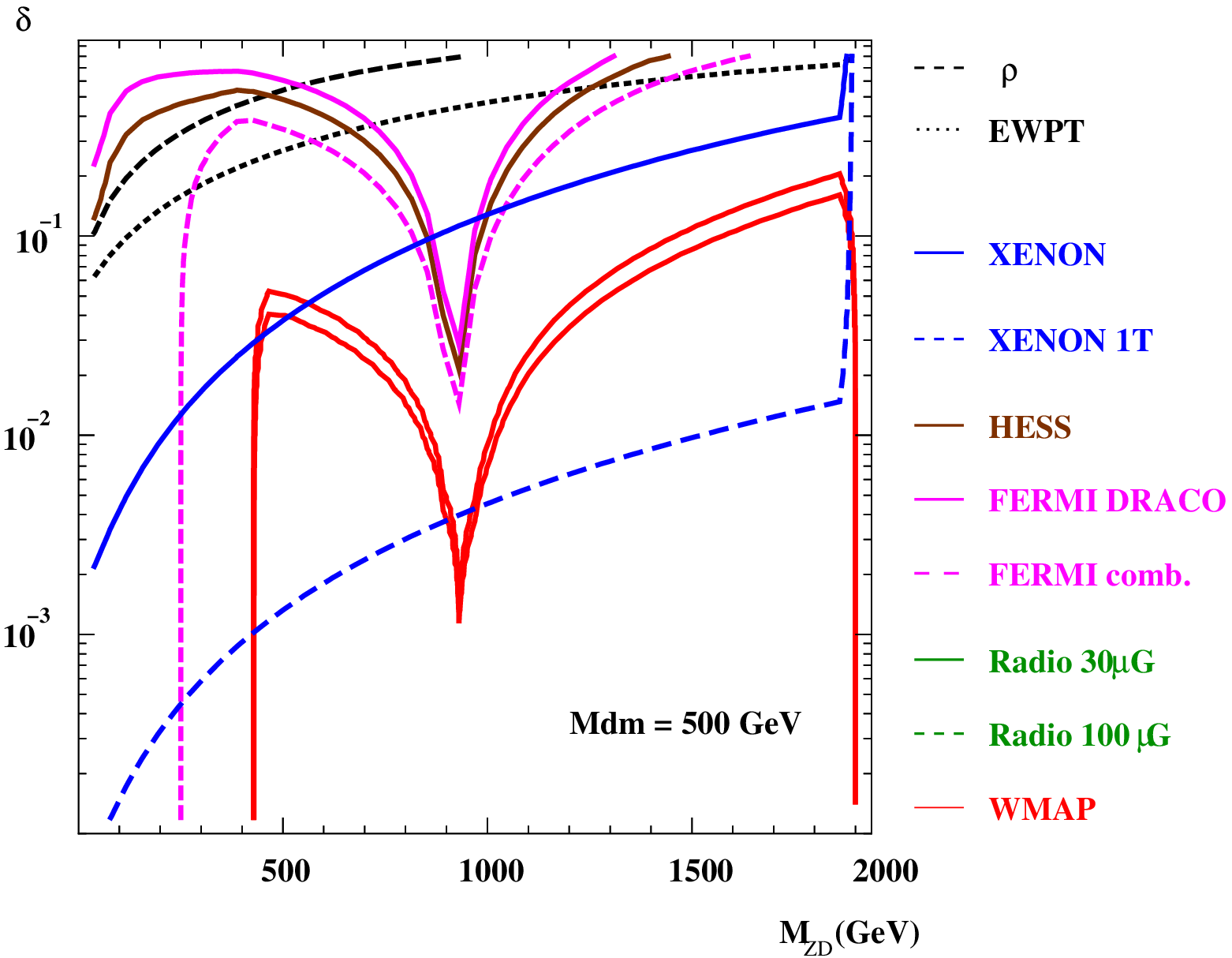}
          \caption{{\footnotesize
Precision test, relic density, direct detection and indirect detection constraints in the
plane ($M_{Z_D}$; $\delta$) for different masses of Dark Matter (5, 25, 200 and 500 GeV)
}}
\label{fig:Scan}
\end{center}
\end{figure}

\section{Constraints}

\noindent

We drove a complete analysis of the allowed parameter space for the kinetic mixing
using the more recent searches in accelerator physics, direct and indirect detection experiments.
Here follows a summary of the constraints we applied in our study:

\subsection{Low energy and electroweak constraints}
Concerning the electroweak symmetry breaking, the mixing between 
$\tilde X_{\mu}$ and $\tilde B_{\mu}$ generates new contributions 
to precision electroweak observables.
However, none of the particle of the SM has any $U_D(1)$ charges:
the $U_D(1)$ can be considered has a $lepto-hadrophobic$ $Z_D$. 
Other authors in \cite{Feldman:2007wj}  or
\cite{Cassel:2009pu} have looked at hidden-valley like models 
or milli--charged dark matter but concentrating their study to relatively
heavy $Z_D$ and a large mixing angle.
 We summarize in this section the main constraints generated 
by such data. 

\begin{itemize}
\item{g-2 :

\noindent
The presence of a light $Z_D$ can contribute to the anomalous magnetic moment of the muon, 
$a_{\mu}=(g_{\mu}-2)/2$. For our analysis we implemented the formulation of Ref.\cite{Fayet:2007ua}. 
To compare with experimental data, we used the latest experimental value \cite{Teubner:2010ah}

\be
\delta a_{\mu}= (31.6 \pm 7.9) \times 10^{-10}
\label{amu}
\ee 

\noindent
We also checked that parity-violation effects \cite{Fayet:2007ua} constrain a much smaller 
region of the parameter space.

\item{$\rho$ parameter :

\noindent
The $\rho$ parameter is defined as $\rho=m_W^2/m_Z^2c_W^2 =1$ in the Standard Model (SM). 
Any electroweak extension of the  SM can generate a deviation to the $\rho$ parameter, especially
if $M_{Z_D} \simeq M_{Z}$, where the mixing is maximum. One can thus translate the global fit of
the $\rho$ parameter $\rho-1= 4^{+8}_{-4} \times 10^{-4}$ \cite{Nakamura:2010zzi} to a constraint
on the ($M_{Z_D}$;$\delta$) plane.
}

\item{
Electroweak precision test :

The authors of \cite{Kumar:2006gm} 
have computed the observables from effective Peskin--Takeuchi 
parameters \cite{EWPT}, and found

\begin{eqnarray}
\Delta m_W &=& (17 \mrm{MeV}) ~\zeta 
\nonumber
\\
\Delta \Gamma_{l+l-} &=& -(8 \mrm{keV})~ \zeta
\nonumber
\\
\Delta \sin^2 \theta_W^{eff} &=& -(0.00033) ~\zeta
\label{Eq:EWPT}
\end{eqnarray}

\noindent
where

\beq
\zeta \equiv 
\left( \frac{\delta}{0.1} \right)^2
\left( \frac{250 \mrm{GeV}}{M_{Z_D}} \right)^2
\eeq

More recently, the authors of 
\cite{Hook} have published an extensive
model independent analysis in the energy range of interest in our study. 
They bounded the kinetic mixing by $\delta \lsim 0.03$ for 
$10 ~ \mrm{GeV}< M_{Z_D} < 200 ~ \mrm{GeV}$ which is in complete agreement
with the constraints given by Eq.(\ref{Eq:EWPT}).

} 
}

\end{itemize}

\subsection{Relic Density}

The abundance of a thermal relic dark matter candidate $\psi_0$ is 
controlled by its annihilation
cross section into SM particles mediated by the exchange of a $Z_D$ gauge
boson through $s-$channel, or $t-$channel $Z_DZ_D$ final state 
(see \cite{Mambrini:2009ad,Dudasline} 
for a detailed study of the relic abundance constraints).
 We modified the micrOMEGAs2.2.CPC code\footnote{The author wants to thank 
 particularly G. Belanger and S. Pukhov for their help to address this issue.}
  \cite{Micromegas} in order to calculate 
 the relic abundance of $\psi_0$ and implemented 
 the points that fulfill the WMAP $5\sigma$ bound \cite{WMAP} ,
 $\Omega_{DM} h^2=0.1123 \pm 5 \times 0.0035$.
One important point is that for a given $M_{Z_D}$ and $m_{\psi_0}$, 
there exists a unique solution $\delta$ (up to the very small uncertainties at 5$\sigma$)
fulfilling WMAP constraints : from 3 parameters
($m_{\psi_0}, M_{Z_D}, \delta$), the WMAP constraints reduce it to two 
($M_{Z_D}, \delta$). 

\noindent
One want to stress that a zone with low relic abundance respecting WMAP is
expected around the region where $M_{Z_{D}} \simeq M_{Z}$. This comes from the 
increase of the  mixing angle $\sin \phi$ in this region of parameter space.
Indeed, for a given $\delta$, $\sin \phi$ increases by a factor of 10 when 
$M_{Z_{D}}$ varies from 90 to 110 GeV (Eq.(\ref{sphi})), decreasing the relic abundance by a factor 100
(the $Z_{D}\bar f f$ coupling is proportional to $\sin \phi$).

\subsection{Direct detection}
There was recently a huge interest concerning the exclusion (or not) of the CoGENT/DAMA/CRESST signal
region corresponding to light dark matter \cite{DAMA,COGENT,X,CRESST}. Several authors tried to re-conciliate the signal
with minimal--like extensions of the standard model : scalar dark matter, extra U(1), ..
Very recently, XENON experiment released a likelihood approach to the first results from XENON100
\cite{Aprile:2011hx}. We took their 
2$\sigma$ limits on their spin--independent elastic WIMP--nucleon cross section. The XENON collaboration also 
recalculated the limit from CDMS assuming an escape velocity of 544 km/s which fits perfectly inside the 2$\sigma$
limits they obtained, as the less conservative scenario on ${\cal L}_{eff}$.  We also explored the possibility of a projected
1 ton XENON experiment \cite{XENON1T}
(more or less equivalent to a super CDMS extension of CDMS \cite{SuperCDMS}).
  During the writing if the article, the XENON experiment published a more detailed analysis
  for light DM exclusion limit with XENON10 data in \cite{Angle:2011th} and with XENON100
  data \cite{Aprile:2011hi}.
  Their results are compatible with the exclusion limit we used through the analysis.

\subsection{Indirect detection}

From the indirect detection data, we combined the very last analysis of different experimental groups.

\begin{itemize}
\item{Observation of Milky Way dwarf spheroidal galaxies by FERMI :
dwarf spheroidal galaxies (dSphs), the largest galactic substructures predicted by the CDM scenario,
are ideal laboratories for indirect searches for dark matter for several reasons: the mass-to-light ratios
in dSphs can be of order 100-1000 showing that they are largely dark matter dominated systems. In 
addition, dSphs are expected to be relatively free from $\gamma$-ray emission from other astrophysical
sources as they have no detected neutral or ionized gas, and little or no recent star formation
activity which make them very interesting objects for indirect dark matter searches, especially with the FERMI
telescope. For our purpose, we will use the two last analysis of the FERMI collaboration: firstly, the study from
the observations of 14 dwarf spheroidal galaxies published in  \cite{Abdo:2010ex}, where we will use the 
data from DRACO observation
as the representative  milky way satellite, with an assumed Navarro-Frenk-White (NFW \cite{NFW})
dark matter density
profile; secondly, as it has been shown in \cite{Garde:2011wr}, it is possible to increase the efficiency of the
constraints using a combined analysis. Even if preliminary, these results are quite robust\footnote{Private communication
with the author of \cite{Garde:2011wr}}.
It is useful to stress that for any indirect detection searches,  experimental analysis are obtained for a given final state.
As we wanted to be the more conservative, we considered their analysis with a $b\overline{b}$ final state.
Indeed, this spectrum is representative of quark spectrum which is the dominant one (due to the color factor)
in the decay of a $Z_{D}$ in $s$ or $t$ channel. Moreover, the final states fraction containing $\mu^+ \mu^-$
or $\tau^+ \tau^-$ (never more than 5 percents in all our points respecting WMAP constraint) produces
a hard $\gamma$-ray spectrum resulting in much more stringent constraints since they predict abundant
photons fluxes at larger energies, where the diffuse background is lower (see \cite{Bernal:2008zk,Goudelis:2009zz}
for an analysis of these specific channels).
For a quark final state, the resulting  rays stem
dominantly from the decay of neutral pions produced in the quark and antiquark hadronization chains, and
do not crucially depend upon the specific quark flavor or mass; in fact, a very similar 
 ray spectrum is
produced by the (typically loop-suppressed) gluon-gluon final state.
 We thus stay conservative in our analysis.
 One can also read \cite{Conrad:2011na} for a clear review of the status and perspectives
 of the FERMI satellite.
}
\item{Searches from the Galactic Center with HESS:
we also included in our analysis the very last searches for a very high $\gamma$-ray signal from
DM annihilation in our galaxy in the region of 45 pc-150 pc after 112h of observations, 
excluding the Galactic plane done
by the HESS collaboration \cite{Abramowski:2011hc}. These limits are among the best reported so far
in this energy range. In this region of the halo, the dependance on the DM profile is quite low, and the $reflected$ 
$background$ technique allows for an intelligent subtraction of the background.
The limits can shift by 30\% due to both the uncertainty on the absolute flux measurement
and the uncertainty of 15\% on the energy scale, which does not affect our general conclusion.
Note that to be consistent with the FERMI analysis, we adopted the NFW profile 
parameterization of the HESS results. }

\item{Radio constraints : charged particles production in a halo is necessarily accompanied by radio waves. 
These signals are produced via the various energy loss processes that charged particles undergo, examples of
which include synchrotron radiation or inverse Compton scattering of electrons from interstellar radiation field
\cite{Crocker:2010gy,Boehm:2010kg,Borriello:2008gy,Ishiwata:2008qy}.
We applied the combined constraints derived using all energy loss processes, featuring only mild dependance
on the astrophysical parameters. Radio data provide more severe restrictions for strong magnetic field $B$. Taking into
account the very large uncertainty on the value of the field in the Galactic center, we studied the constraints
for $B=30 \mu$G and 100$\mu$G, with the NFW profile to stay consistent with the HESS and FERMI searches. We
followed the recent analysis given in \cite{Crocker:2010gy}. }

\item{
 We also implemented  the constraints derived from FERMI diffuse measurements  
\cite{Zaharijas:2010ca}. However, these constraints were always lying between the constraints coming from 
the DRACO satellite and the ones from the combined dSphs analysis. We thus decided  not to show them in the plots.}

\end{itemize}

\section{Results}

\subsection{Where does hit each experiment?}

We show in Fig.\ref{fig:Scan} the electroweak, cosmological and astrophysical constraints in the plane
($M_{Z_D}$;$\delta$) for different DM masses (5, 25, 200 and 500 GeV). One  can distinguish 4 regimes :
very light DM ($\sim 5$ GeV) light DM ($\sim 30$ GeV), heavy ($\sim 100$ GeV) and very heavy 
($\gtrsim 500$ GeV) DM. Depending on the regime, some constraints will dominate over other ones, 
making complementarity between all these detections mode fundamental for a general analysis of any
$U'(1)$ extension of the SM.
Another interesting region is the parameter space with maximal mixing, where $M_{Z_D} \sim M_Z$. However,
this region when respecting WMAP constraints is totally excluded because of a large excess of the $\rho$
parameter.

\subsection{The (very)light Dark Matter regime}

In the case of very light dark matter ($\lesssim 10$ GeV), the XENON experiment is relatively inefficient.
As one can check in Fig.\ref{fig:Scan} (top) all the points respecting WMAP respect also the last data from 
XENON100. However, some other experiments like FERMI and especially radio constraints
become more efficient as was already stressed by the authors of \cite{Boehm:2010kg}
where they showed that a DM mass between 
1 and 10 GeV could be excluded by radio data from galactic center. One also notices that the electroweak 
constraints become the more effective constraints for masses of $Z_D$, $M_{Z_D}\gtrsim 55$ GeV where
they exclude the region respecting WMAP. Indeed, these
constraints are obviously independent of the DM mass, but are stringent when the $Z-Z_D$ mixing is important
($M_{Z_D}\sim M_{Z}$).

\noindent
For $25 \mrm{GeV} \lesssim m_{\psi_0} \lesssim 200$ GeV, one can see that the radio constraints
 become less effective, whereas
the XENON experiment excludes a large part of the parameter space allowed by WMAP except near the pole.
 Indeed
in the region ($M_{Z_D}$;$\delta$) where $2 m_{\psi_0} \simeq M_{Z_D}$, the $s$-channel $Z_D$ exchange cross section
increases dramatically. Points respecting WMAP lie in region of parameter space with low values of $\delta$ 
(the "sea-gull" shape in Figs.\ref{fig:Scan}) : with such low values of $\delta$, the direct detection (through $t$-channel
$Z_D$ exchange) cross section $\sigma_{p \psi_0}$ stays below the XENON100 exclusion limit. However, this
$Z_D-$pole region is sensitive to indirect detection experiment like FERMI or HESS because of the high value
of the annihilating rate $\psi_0 \psi_0 \rightarrow Z_D \rightarrow \bar f f$.

\subsection{The (very)heavy Dark Matter regime}

In this regime, a XENON--like experiment becomes less effective (because of the mass suppression in the scattering
cross section) whereas FERMI and HESS indirect detection data exclude the larger part of the parameter space around
the pole.
However, one can check in Fig.\ref{fig:Scan}(bottom) that a 500 GeV DM is not yet excluded, except in the region where 
$M_{Z_D}\lesssim m_{\psi_0}$, dominated by the $\psi_0 \psi_0 \rightarrow Z_D Z_D$ annihilation final state.
HESS limits become more efficient than FERMI for $m_{\psi_0}\gtrsim 300$ GeV.
Contrary to the (very)light case, the electroweak observables are always less constraining than any astrophysical
data given by direct or indirect detection.
Whatever is the DM mass, one can check in Figs.\ref{fig:Scan}(all)
that a XENON 1 ton experiment would be able to exclude all the region 
respecting WMAP except in a very narrow region around the pole if $m_{\psi_0}\gtrsim 500$ GeV.


\begin{figure}
    \begin{center}
    \includegraphics[width=1.5in]{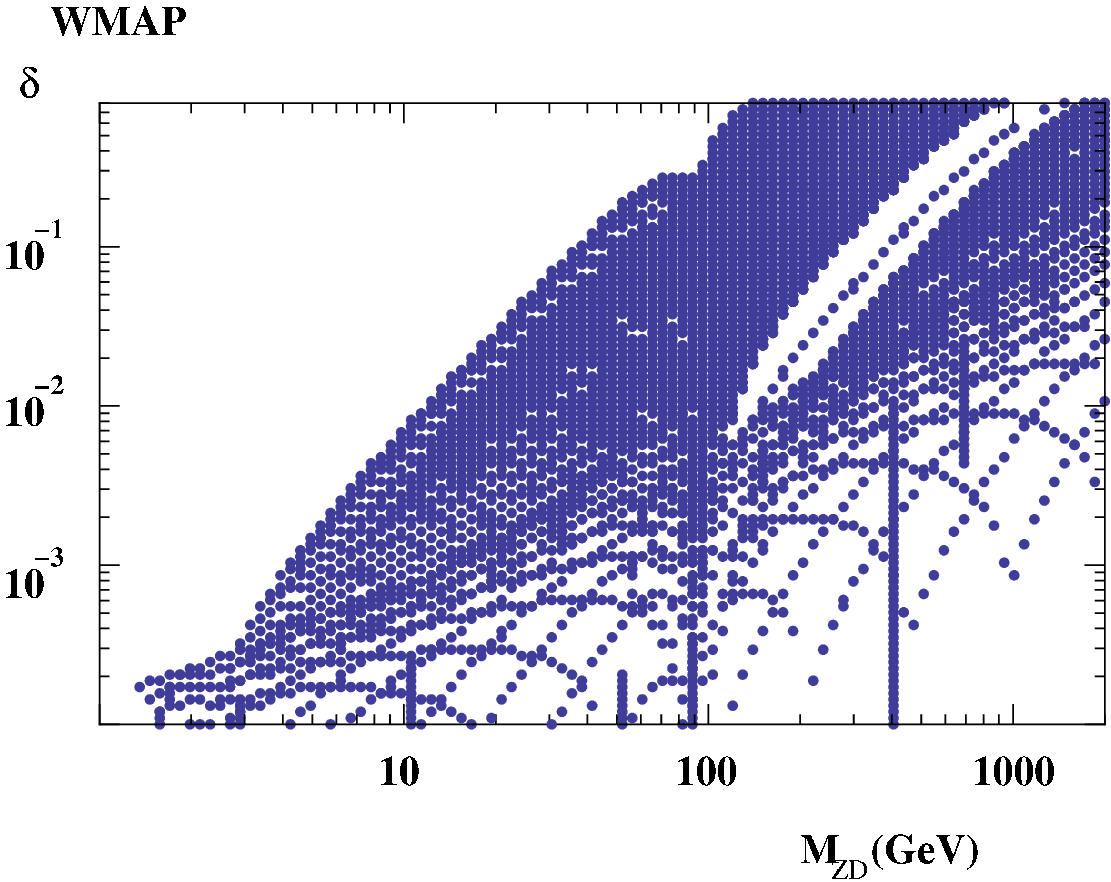}
    \includegraphics[width=1.5in]{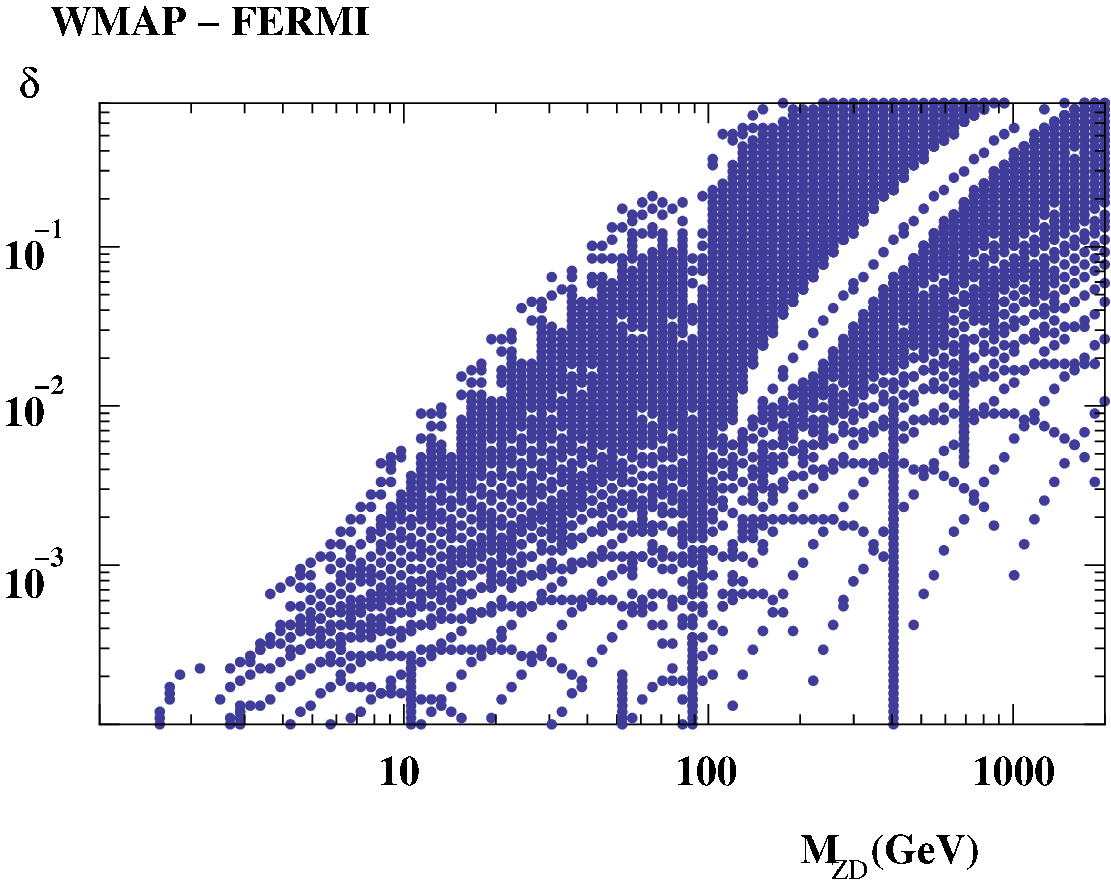}
    
        \includegraphics[width=1.5in]{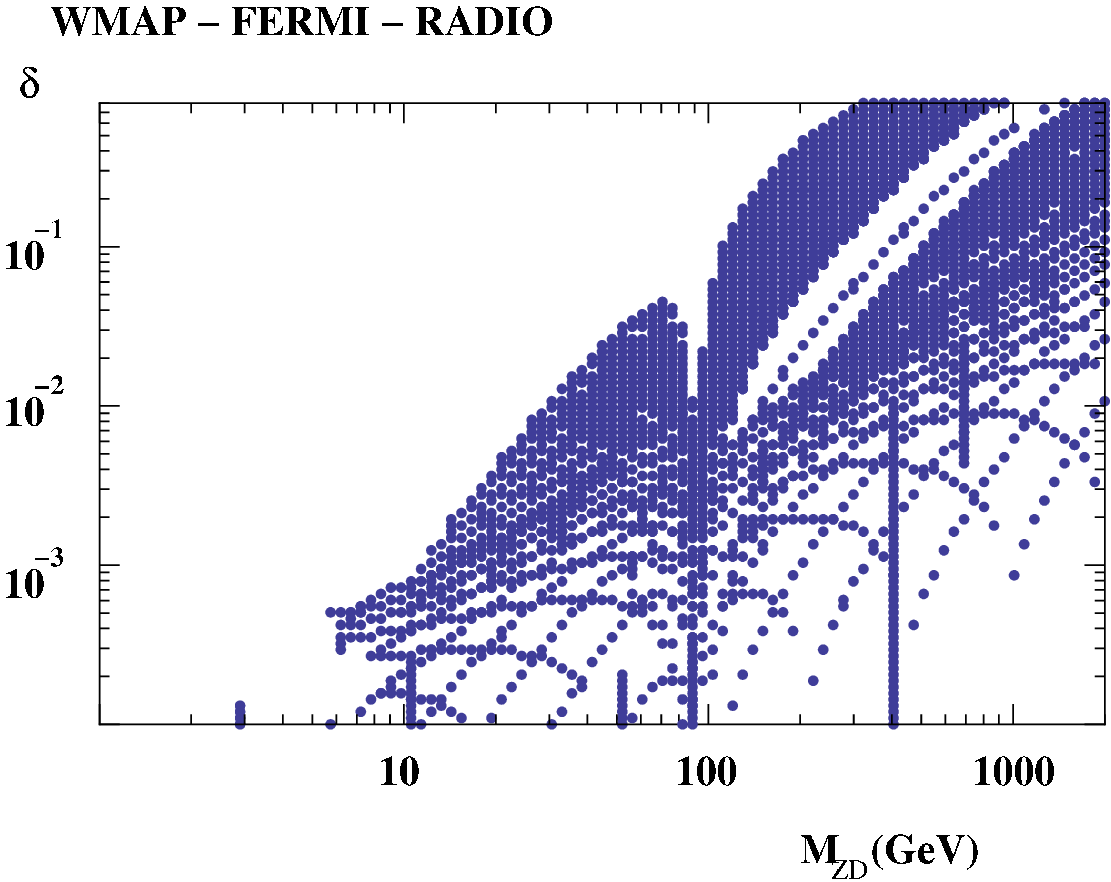}
    \includegraphics[width=1.5in]{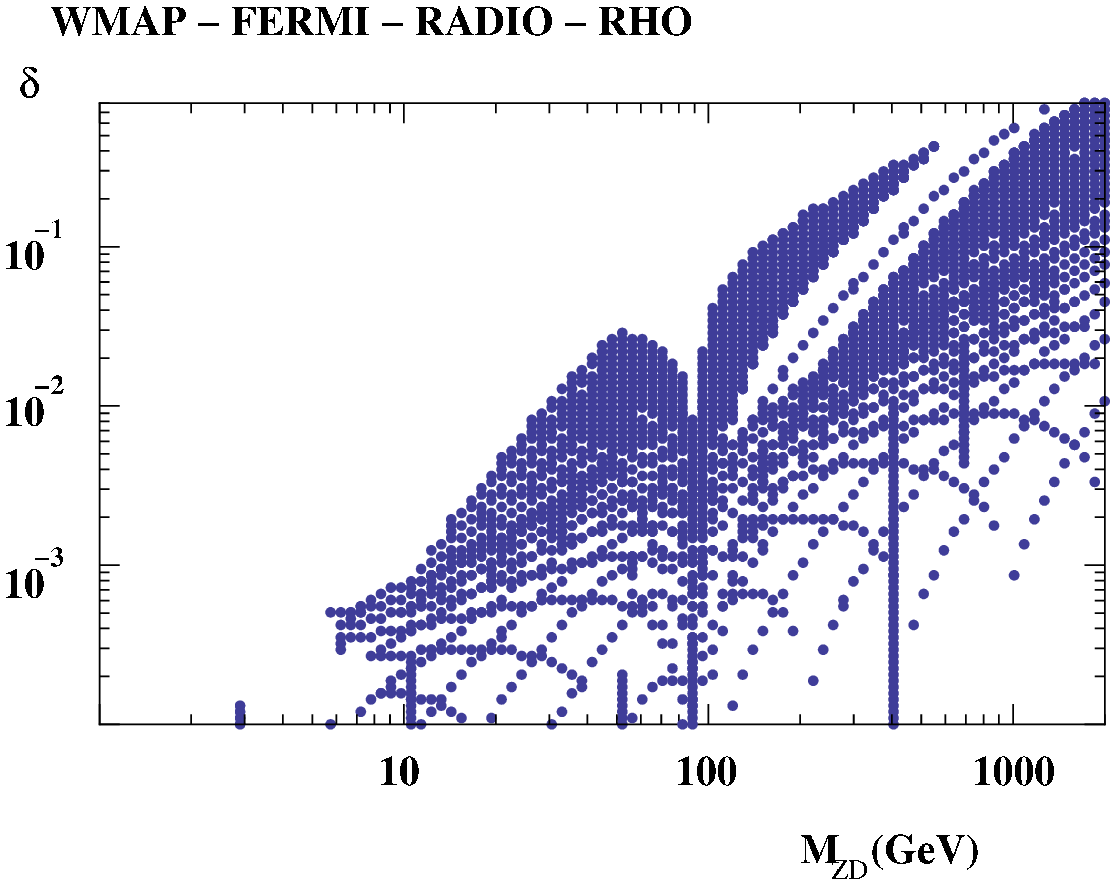}    
    
    \includegraphics[width=1.5in]{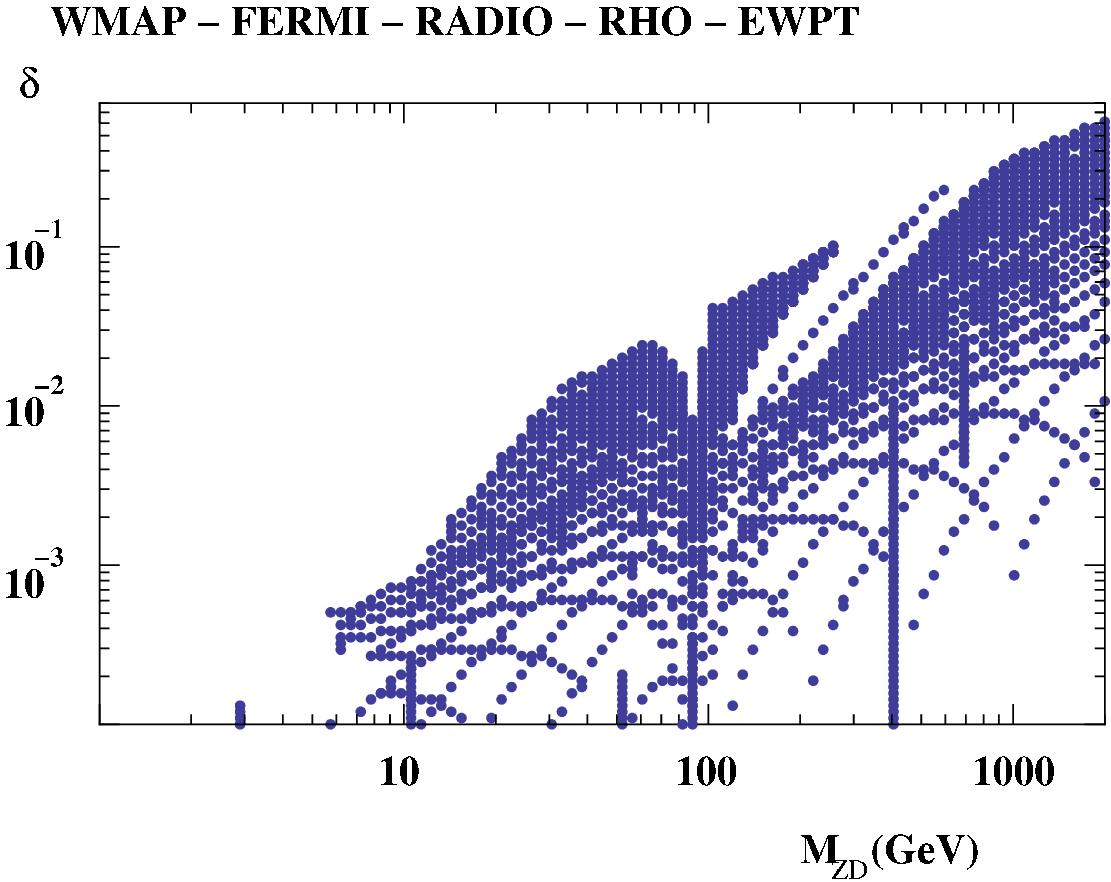}
    \includegraphics[width=1.5in]{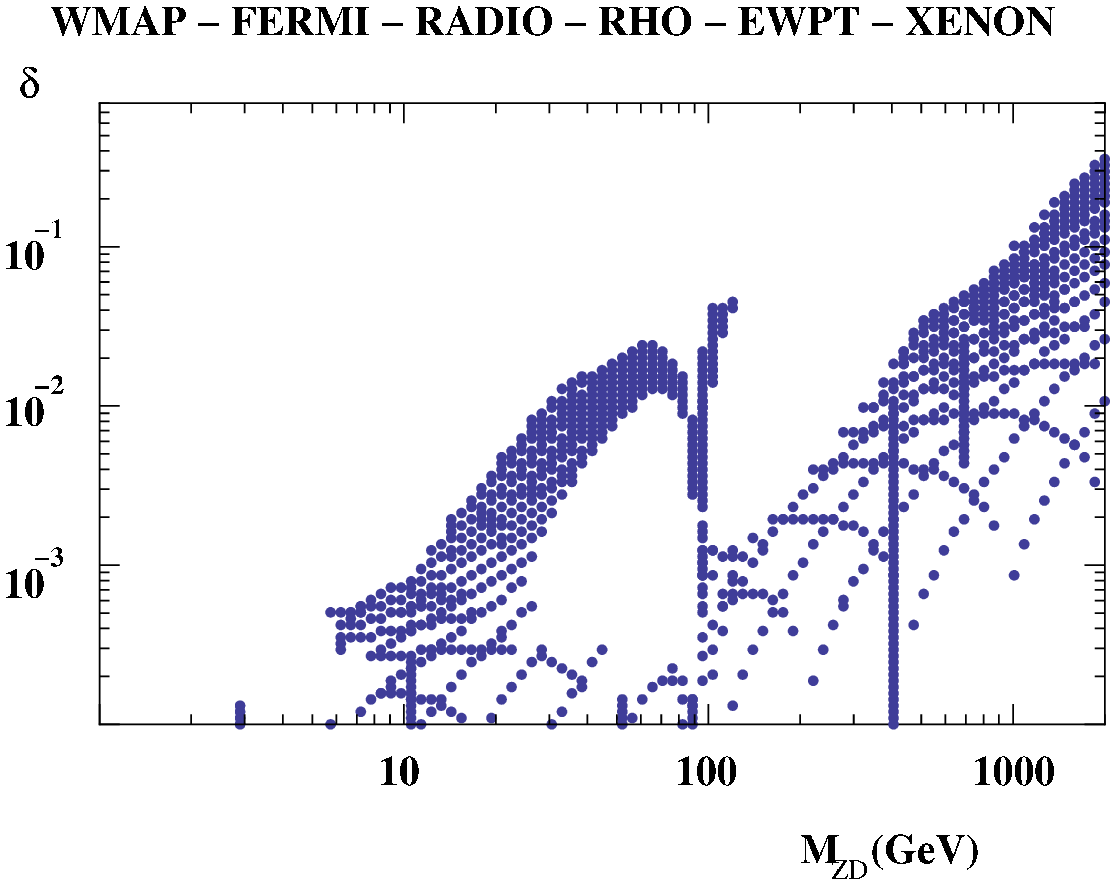}
    
        \includegraphics[width=1.5in]{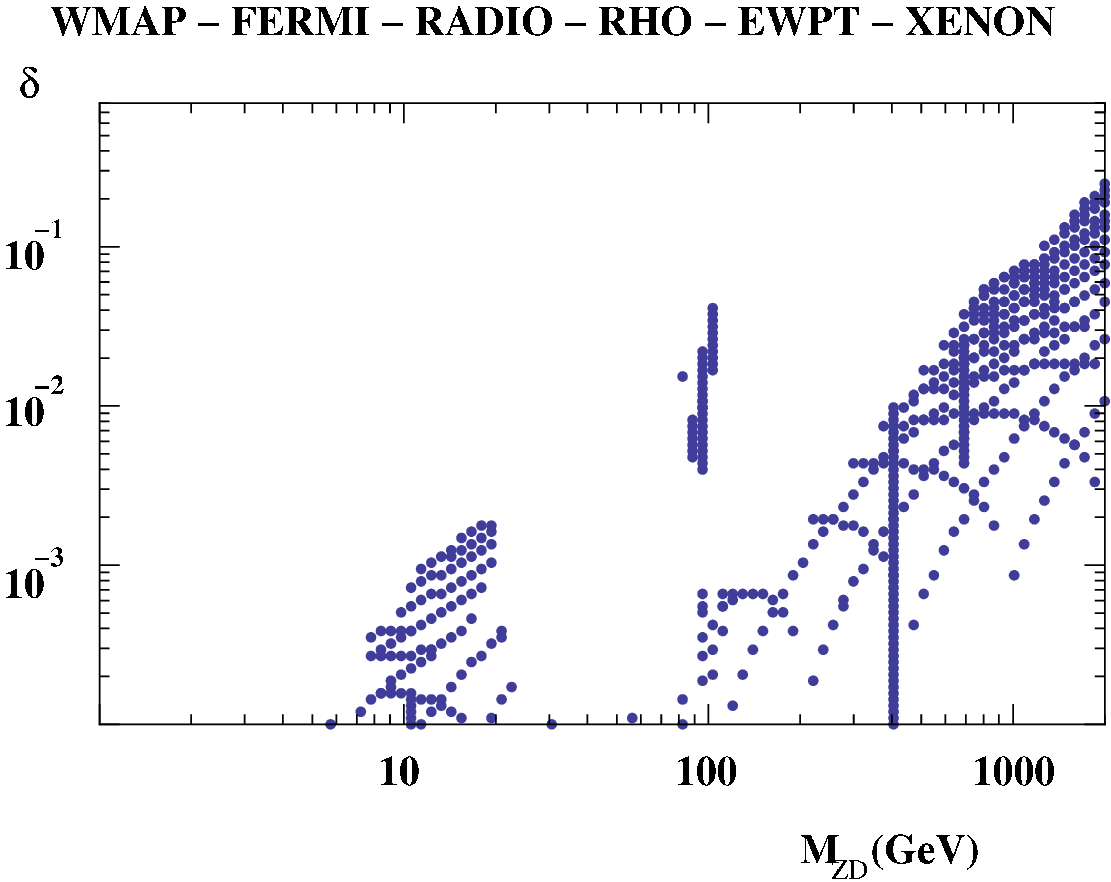}
    \includegraphics[width=1.5in]{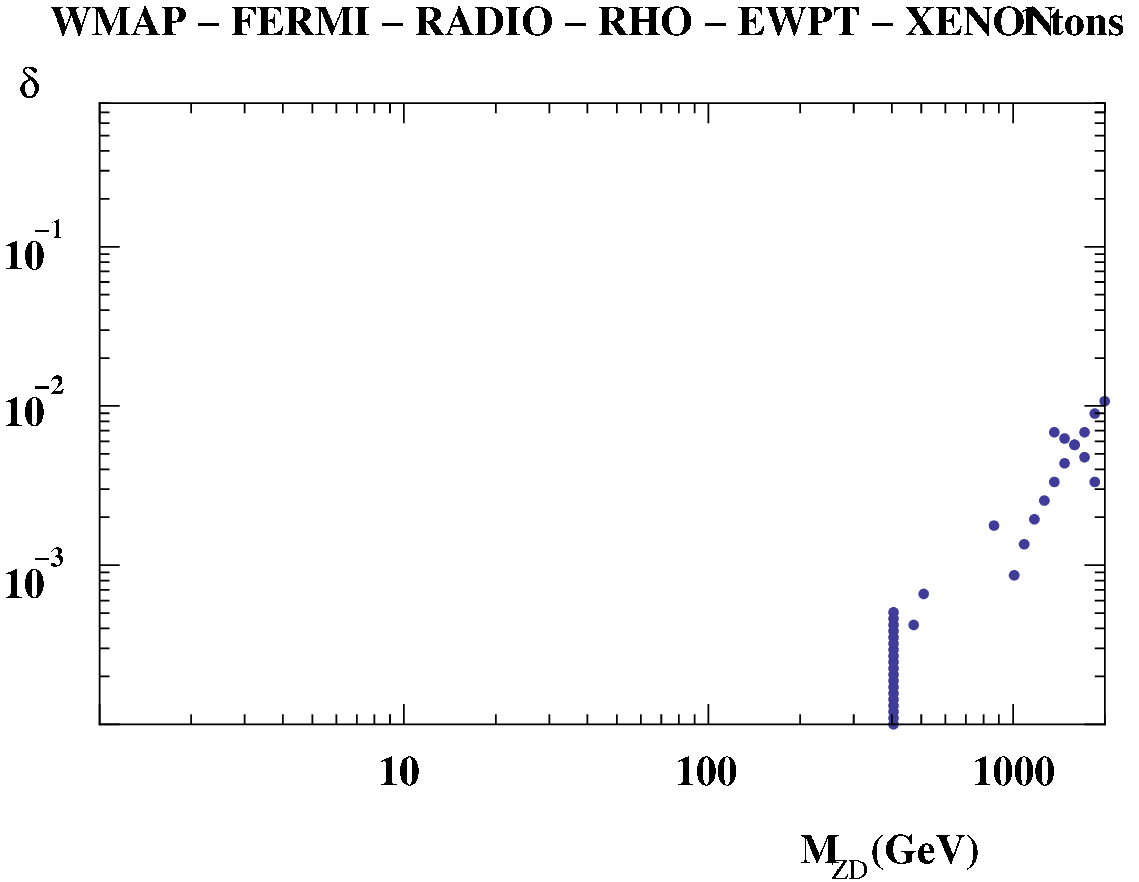}

          \caption{{\footnotesize
Parameter space allowed taking account different type of experiments in the ($m_{\psi_0}$;$M_{Z_D}$) plane.
The two plots taking into account the XENON constraints correspond to the $+ 2 \sigma$ and $-2 \sigma$ zones
of exclusion.
}}
\label{fig:Summary}
\end{center}
\end{figure}

We also analyzed the allowed region in the ($M_{Z_D}$;$\delta$) parameter space after a complete scan on the
DM mass $1 \mrm{GeV} < m_{\psi_0}< 1$ TeV, taking into account the different natures of experimental exclusions.
 The results are presented
in Figs.\ref{fig:Summary}. With such a scan, our results become independent of the nature, coupling or mass
of the dark matter candidate, and one can compare directly the sensitivity of each type of experiment for
specific  regions of the parameter space.

It appears clearly that FERMI experiment excludes a region of the parameter space corresponding
to $M_{ZD} \lesssim 100$ GeV which is the region where the satellite is the more effective. The radio constraints
exclude a part of the parameter with lighter $Z_D$ as expected \cite{Crocker:2010gy,Boehm:2010kg}.
 The $\rho$ and EWPT constraints reach the region
of the parameter space where $M_{Z_D}\simeq M_Z$ (maximal mixing). Even after taking into account the XENON100 
exclusion region at $\pm 2 \sigma$, a region respecting WMAP survives around $M_{Z_D} \simeq 10$ GeV where 
one can fit COGENT, CRESST or DAMA excesses \cite{Mambrini:2010dq,Mambrini:2010yp} .
One also observes that a 1 ton XENON--like experiment would exclude 99 \% of the parameter space allowed 
by WMAP. It would be quite difficult to escape such constraints as the couplings of the $Z_D$ to the quarks
are fixed for a given value of $\delta$ .

We would like to say few words on the "hole" one observes in the parameter space 
-before taking into account the XENON100 constraints- in Figs.\ref{fig:Summary}(top). 
For a given $M_{Z_D}$, when $\delta$ increases, the $Z_D$ pole "disappears" in the sense
that the relic abundance stays below WMAP data between the $Z$ and $Z_D$-pole: 
points  that respect WMAP constraints lie $very$ near from the $Z$-pole, especially because for such large
values of $\delta$, the $Z-$pole branch is extremely narrow. Indeed, $Z\psi_0\psi_0$ coupling is proportional to
$\sin\phi$, Eq.(\ref{sphi}), which means that the annihilation cross section 
$\langle \sigma v \rangle \propto \delta^2/ \Gamma_Z^2$ ($\Gamma_Z$ being the width of the $Z$ boson)
is very sensitive.
When $\delta$ increases from $10^{-3}$ to $10^{-1}$, the cross section gains 4 orders of magnitude!
One needs an extremely fine tuning (at the 0.1\% level) on the condition $m_{\psi_0}=M_Z/2$ to respect WMAP bounds. 
This extreme fine-tuning condition
respects a law $\sin \phi$=cste $\Rightarrow \delta \propto M_{Z_D}$ (for $M_{Z_D} \gtrsim M_Z$, 
Eq.(\ref{sphi})) which is exactly
the shape of the hole of Fig.\ref{fig:Summary}(top). For much higher values of $\delta$,
points do not need to be on the $Z-$pole anymore to respect WMAP (for $\delta=0.8$ and $M_{Z_D}=300$ GeV,
 one enters in WMAP bounds for $m_{\psi_0}\simeq 3$ GeV).
 In any case, this region of extreme fine-tuning is largely excluded by XENON100 
 experiment, even in the more conservative case. 

\begin{figure}
    \begin{center}
    \hspace{-0.8cm}
    \includegraphics[width=1.7in]{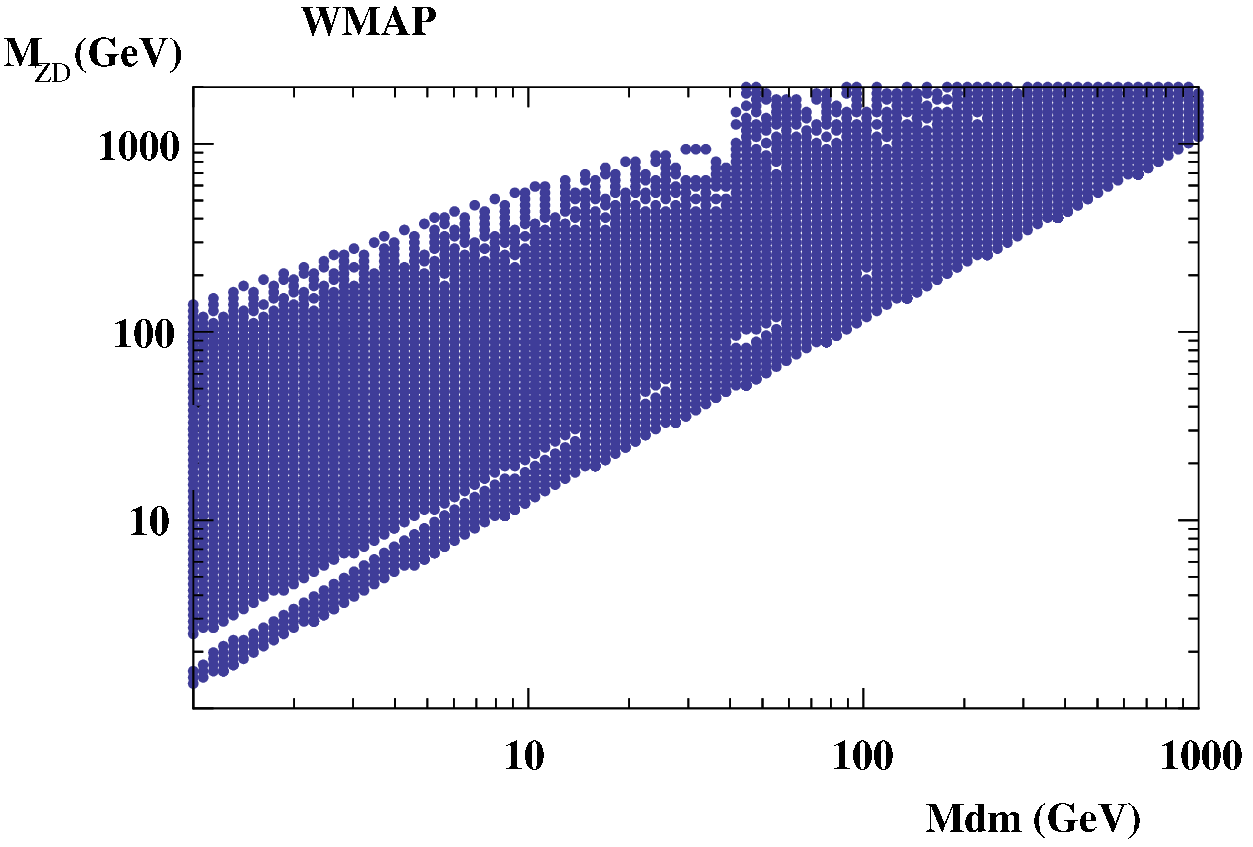}
    \includegraphics[width=1.7in]{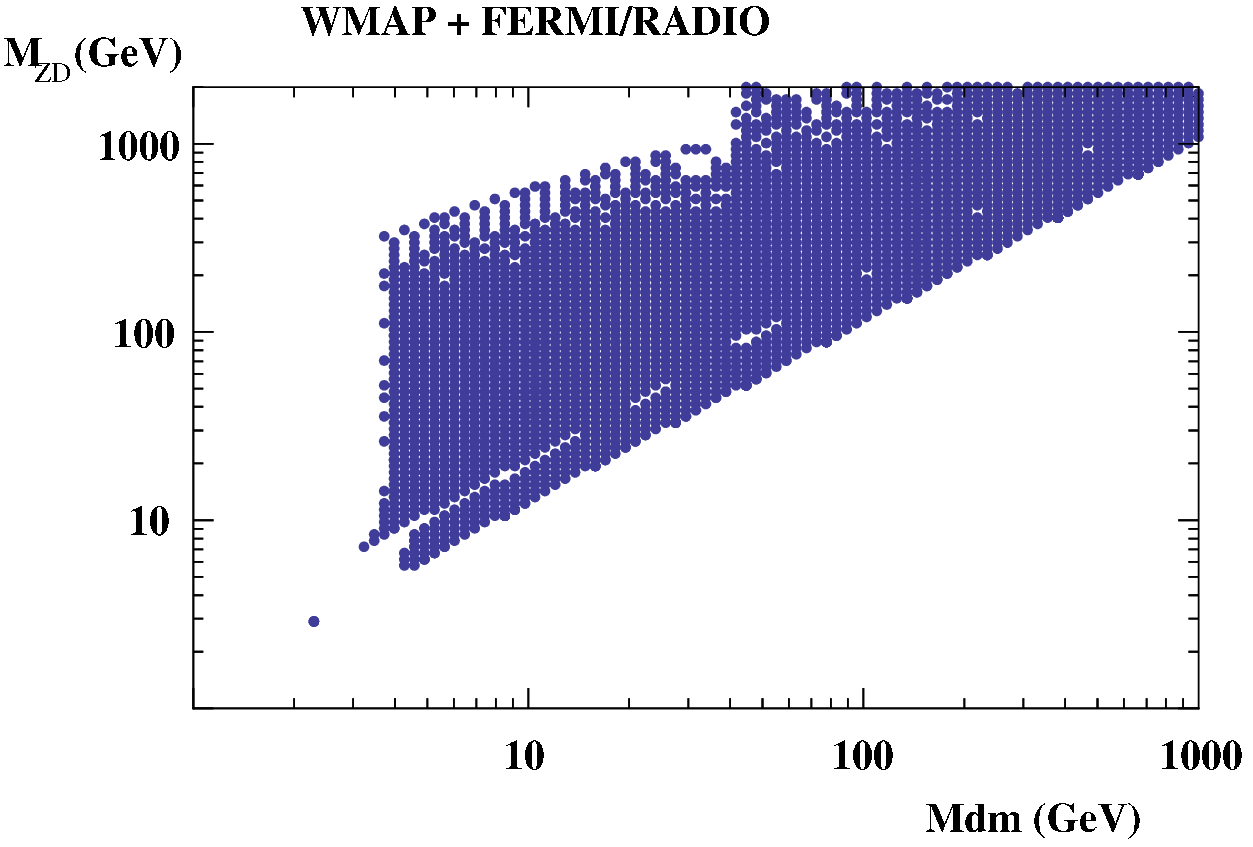}
    
    \hspace{-0.8cm}
        \includegraphics[width=1.7in]{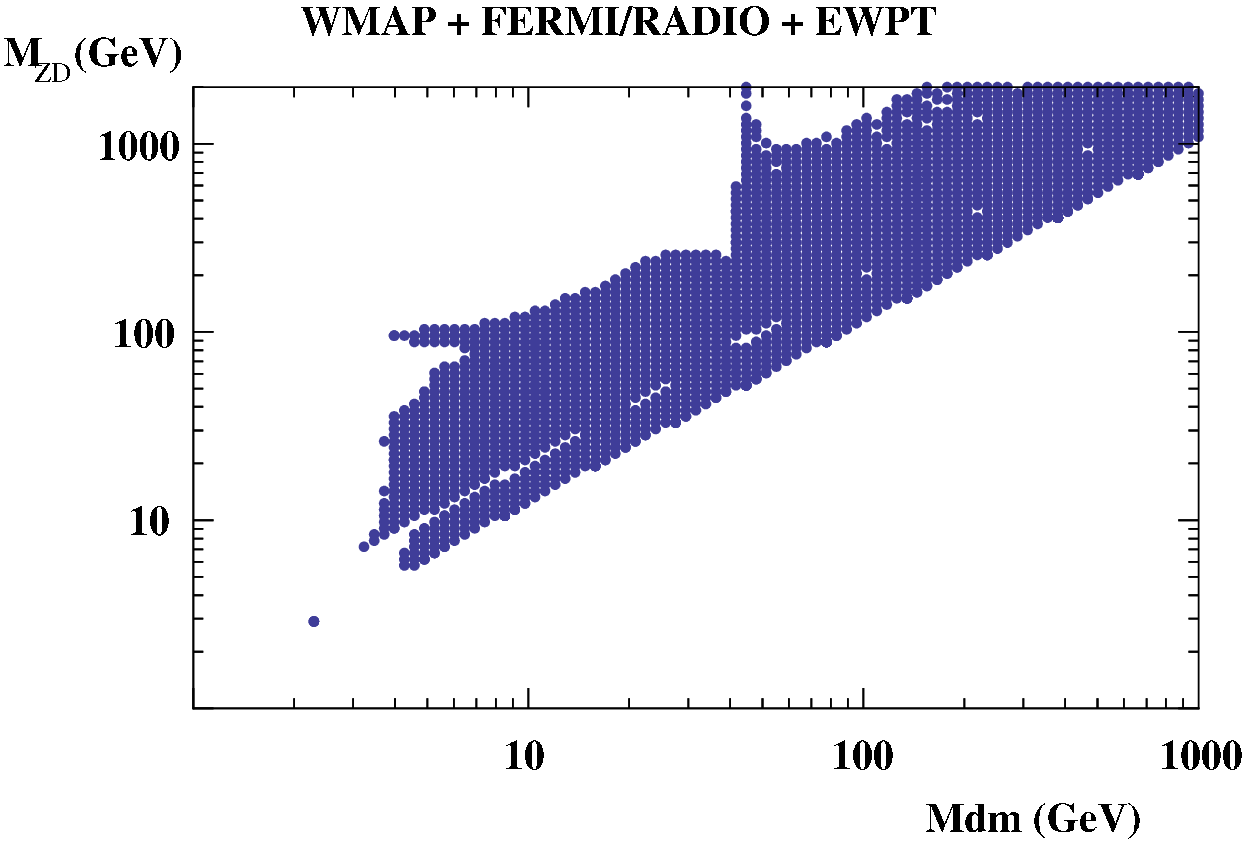}
    \includegraphics[width=1.7in]{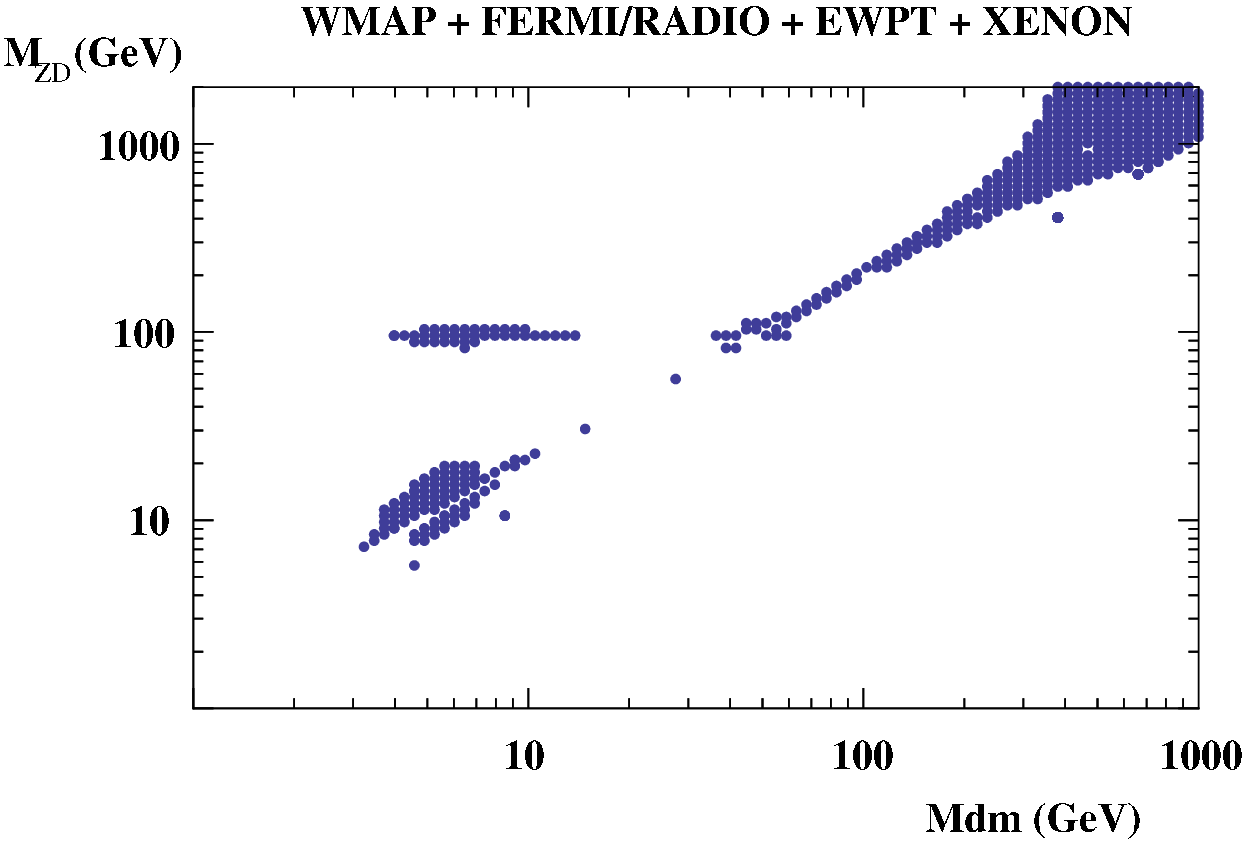}    
    
          \caption{{\footnotesize
Parameter space allowed within 90 \% of  C.L. for the CoGeNT
signal (blue), DAMA without channeling (red), with channeling 
(green), CRESST (black), and the exclusion region depending on the hypothesis
concerning $L_{eff}$ (se the text for details).
}}
\label{fig:Mpsi}
\end{center}
\end{figure}

\subsection{Consequences on $m_{\psi_0}$ and $M_{Z_D}$}

One interesting analysis was to look at the consequences on $m_{\psi_0}$ independently of 
$\delta$. 
We show in Figs.\ref{fig:Mpsi} the region of the parameter space allowed by the different type of experiments
in the ($m_{\psi_0}$;$M_{Z_D}$) plane after a scan on $10^{-4} < \delta < 0.8$. One observes that WMAP
restricts the parameter space to a relatively broad region around the line $M_{Z_D} \sim 2 m_{\psi_0}$.
After taking into account the indirect detection constraints, the region of light mass is excluded as expected. The EWPT
and $\rho$-parameter exclude the region  near $M_{Z_D} \sim M_{Z}$ (maximal mixing) and
further from the pole because the further one stays from the pole region, the higher the value of 
$\delta$ should be to respect WMAP\footnote{We also notice that even more stringent constraints 
from WMAP could have been applied \cite{Hutsi:2011vx}, but we wanted to stay as conservative 
as possible in our study. }, at the risk to be excluded by EWPT.
Once we consider the exclusion limit by XENON100, the only points surviving are the ones lying exactly on the pole line
$M_{Z_D}=2m_{\psi_0}$: in this region, one needs on the contrary a very small value of $\delta$ ($\sim 10^{-3}$)
to respect WMAP because of the high annihilation cross section of the process 
$\psi_0 \psi_0 \rightarrow  Z_D \rightarrow \bar f f$. Such small values for $\delta$ respect XENON100
exclusion limits.
One also notices a vertical breaking step in the parameter space
for $m_{\psi_0}\simeq 45$ GeV (before applying the XENON100 constraints) which corresponds
to the $Z-$pole region. 
It is interesting to point out that there still exists a bulk region around $m_{\psi_0} \simeq 7$ GeV not yet excluded by
XENON and which is able to explain the last CRESST/CoGENT data \cite{Mambrini:2010dq,Hooper:2010uy} . 
We should mention that this model has a common feature with the singlet extension of the SM studied
in \cite{Tytgat, Tytgatbis} : the same diagram which gives the relic abundance is the one contributing to the
nucleon-scattering (from $s-$ to $t-$ channel $Z_D$ exchange). One can demonstrate that 
these sorts of models usually ensure a large direct detection rate.

\section{Conclusion and Summary}

We showed that the existence of a $dark$
$U_D(1)$ gauge sector which interacts with the Standard Model only through 
its kinetic mixing possesses a valid dark matter candidate respecting
accelerator, cosmological and the more recent direct detection constraints.
We observed that when one combines direct and indirect detection searches, the
restrictions are much more stringent that the ones extracted by precision test measurements.
All these experiments are indeed complementary, excluding different parts of the
parameter space allowed by WMAP.
Moreover, once we take into account all these constraints, only a narrow region of the parameter
space along the $Z_D$-pole region $M_{Z_D}= 2 m_{\psi_0}$ survives. There is still a  bulk region
around $m_{\psi_0}\simeq 7$ GeV which can explain the recent excesses observed in some
direct detection experiments. We also showed that  a 1ton XENON--like experiment
would be able to test more than 90\% of the remaining parameter space, being able to exclude (discover?)
robustly  such models.


\section*{Acknowledgements}
Y.M. wants to thank particularly E. Dudas, G. Belanger,
 A. Romagnoni and Bryan Zald\'\i var for useful discussions. The work was
supported by the french ANR TAPDMS {\bf ANR-09-JCJC-0146} 
and the spanish MICINNÕs Consolider-Ingenio 2010 Programme 
under grant  Multi- Dark {\bf CSD2009-00064}.



\begin{thebibliography}{99}







\bibitem{Langacker:2008yv}
  P.~Langacker,
  ``The Physics of Heavy $Z^\prime$ Gauge Bosons,''
  Rev.\ Mod.\ Phys.\  {\bf 81} (2008) 1199
  [arXiv:0801.1345 [hep-ph]].





\bibitem{Holdom}
  R.~Foot, X.~-G.~He,
  ``Comment on Z Z-prime mixing in extended gauge theories,''
  Phys.\ Lett.\  {\bf B267 } (1991)  509-512;
    R.~Foot, H.~Lew, R.~R.~Volkas,
  ``A Model with fundamental improper space-time symmetries,''
  Phys.\ Lett.\  {\bf B272 } (1991)  67-70;
  B.~Holdom,
  ``Two U(1)'S And Epsilon Charge Shifts,''
  Phys.\ Lett.\  B {\bf 166}, 196 (1986).

\bibitem{Feldman:2007wj}
  D.~Feldman, Z.~Liu and P.~Nath,
  ``The Stueckelberg Z' extension with kinetic mixing and milli-charged dark
  matter from the hidden sector,''
  Phys.\ Rev.\  D {\bf 75} (2007) 115001
  [arXiv:hep-ph/0702123].


\bibitem{Martin:1996kn}
  S.~P.~Martin,
  ``Implications of supersymmetric models with natural R-parity conservation,''
  Phys.\ Rev.\  D {\bf 54} (1996) 2340
  [arXiv:hep-ph/9602349].

\bibitem{Rizzo:1998ut}
  T.~G.~Rizzo,
  ``Gauge kinetic mixing and leptophobic $Z^\prime$ in E(6) and SO(10),''
  Phys.\ Rev.\  D {\bf 59} (1999) 015020
  [arXiv:hep-ph/9806397].


\bibitem{delAguila:1995rb}
  F.~del Aguila, M.~Masip and M.~Perez-Victoria,
  ``Physical parameters and renormalization of U(1)-a x U(1)-b models,''
  Nucl.\ Phys.\  B {\bf 456} (1995) 531
  [arXiv:hep-ph/9507455].

\bibitem{Dobrescu:2004wz}
  B.~A.~Dobrescu,
  ``Massless gauge bosons other than the photon,''
  Phys.\ Rev.\ Lett.\  {\bf 94} (2005) 151802
  [arXiv:hep-ph/0411004].

\bibitem{Dienes:1996zr}
  K.~R.~Dienes, C.~F.~Kolda and J.~March-Russell,
  ``Kinetic mixing and the supersymmetric gauge hierarchy,''
  Nucl.\ Phys.\  B {\bf 492} (1997) 104
  [arXiv:hep-ph/9610479].



\bibitem{Cohen:2010kn}
  T.~Cohen, D.~J.~Phalen, A.~Pierce and K.~M.~Zurek,
  ``Asymmetric Dark Matter from a GeV Hidden Sector,''
  arXiv:1005.1655 [hep-ph].


\bibitem{Kang:2010mh}
  Z.~Kang, T.~Li, T.~Liu, C.~Tong, J.~M.~Yang,
  JCAP {\bf 1101 } (2011)  028.
  [arXiv:1008.5243 [hep-ph]].



\bibitem{Feldman:2006wd}
  D.~Feldman, B.~Kors and P.~Nath,
  ``Extra-weakly Interacting Dark Matter,''
  Phys.\ Rev.\  D {\bf 75}, 023503 (2007)
  [arXiv:hep-ph/0610133].







\bibitem{Pospelov:2008zw}
  M.~Pospelov,
  ``Secluded U(1) below the weak scale,''
  Phys.\ Rev.\  D {\bf 80} (2009) 095002
  [arXiv:0811.1030 [hep-ph]];
M.~Pospelov, A.~Ritz and M.~B.~Voloshin,
  ``Secluded WIMP Dark Matter,''
  Phys.\ Lett.\  B {\bf 662} (2008) 53
  [arXiv:0711.4866 [hep-ph]];
  W.~F.~Chang, J.~N.~Ng and J.~M.~S.~Wu,
  ``A Very Narrow Shadow Extra Z-boson at Colliders,''
  Phys.\ Rev.\  D {\bf 74} (2006) 095005
  [Erratum-ibid.\  D {\bf 79} (2009) 039902]
  [arXiv:hep-ph/0608068];
  Z.~Liu,
  ``Hidden Sector Models and Signatures,''
  Nucl.\ Phys.\ Proc.\ Suppl.\  {\bf 200-202} (2010) 133
  [arXiv:0910.0061 [hep-ph]];
   R.~Foot,
  ``A comprehensive analysis of the dark matter direct detection experiments in the mirror dark matter framework,''
  Phys.\ Rev.\  {\bf D82 } (2010)  095001.
  [arXiv:1008.0685 [hep-ph]].













\bibitem{Cicoli:2011yh}
  M.~Cicoli, M.~Goodsell, J.~Jaeckel and A.~Ringwald,
  ``Testing String Vacua in the Lab: From a Hidden CMB to Dark Forces in Flux
  Compactifications,''
  arXiv:1103.3705 [hep-th].

\bibitem{Kumar:2007zza}
  J.~Kumar, A.~Rajaraman and J.~D.~Wells,
  ``Probing the Green-Schwarz Mechanism at the Large Hadron Collider,''
  Phys.\ Rev.\  D {\bf 77} (2008) 066011
  [arXiv:0707.3488 [hep-ph]].

 


\bibitem{Javier}
 M.~Goodsell, J.~Jaeckel, J.~Redondo and A.~Ringwald,
  ``Naturally Light Hidden Photons in LARGE Volume String Compactifications,''
  JHEP {\bf 0911} (2009) 027
  [arXiv:0909.0515 [hep-ph]];
  
 S.~A.~Abel, M.~D.~Goodsell, J.~Jaeckel, V.~V.~Khoze and A.~Ringwald,
  ``Kinetic Mixing of the Photon with Hidden U(1)s in String Phenomenology,''
  JHEP {\bf 0807} (2008) 124
  [arXiv:0803.1449 [hep-ph]];


\bibitem{Cassel:2009pu}
  S.~Cassel, D.~M.~Ghilencea and G.~G.~Ross,
  ``Electroweak and Dark Matter Constraints on a Z' in Models with a Hidden
  Valley,''
  Nucl.\ Phys.\  B {\bf 827} (2010) 256
  [arXiv:0903.1118 [hep-ph]].






\bibitem{Chun:2010ve}
  E.~J.~Chun, J.~C.~Park and S.~Scopel,
  ``Dark matter and a new gauge boson through kinetic mixing,''
  JHEP {\bf 1102} (2011) 100
  [arXiv:1011.3300 [hep-ph]].







\bibitem{Mambrini:2010yp}
  Y.~Mambrini,
  ``Specific Dark Matter signatures from hidden U(1),''
  arXiv:1012.0447 [hep-ph].







\bibitem{CRESST}
 G.~Angloher {\it et al.},
  ``Limits on WIMP dark matter using sapphire cryogenic detectors,''
  Astropart.\ Phys.\  {\bf 18} (2002) 43;
see talk by W. Seidel, WONDER 2010 Workshop, Laboratory Nazionali del Gran Sasso, Italy, March 22-23, 2010 and MPIK seminar by T. Schwetz, june21, 2010.

  \bibitem{COGENT}
  C.~E.~Aalseth {\it et al.}  [CoGeNT collaboration],
  ``Results from a Search for Light-Mass Dark Matter with a P-type Point
  Contact Germanium Detector,''
  arXiv:1002.4703 [astro-ph.CO].




\bibitem{DAMA}
  R.~Bernabei {\it et al.},
  ``Dark matter search,''
  Riv.\ Nuovo Cim.\  {\bf 26N1} (2003) 1
  [arXiv:astro-ph/0307403];
 R.~Bernabei {\it et al.}  [DAMA Collaboration],
  ``First results from DAMA/LIBRA and the combined results with DAMA/NaI,''
  Eur.\ Phys.\ J.\  C {\bf 56} (2008) 333
  [arXiv:0804.2741 [astro-ph]].






\bibitem{Mambrini:2010dq}
  Y.~Mambrini,
  ``The Kinetic dark-mixing in the light of CoGENT and XENON100,''
  JCAP {\bf 1009}, 022 (2010)
  [arXiv:1006.3318 [hep-ph]].
  
\bibitem{Hooper:2010uy}
  D.~Hooper, J.~I.~Collar, J.~Hall and D.~McKinsey,
  ``A Consistent Dark Matter Interpretation For CoGeNT and DAMA/LIBRA,''
  Phys.\ Rev.\  D {\bf 82}, 123509 (2010)
  [arXiv:1007.1005 [hep-ph]];
   M.~Buckley, P.~F.~Perez, D.~Hooper and E.~Neil,
  arXiv:1104.3145 [hep-ph].

  
\bibitem{Cheung}
  K.~Cheung, K.~H.~Tsao and T.~C.~Yuan,
  ``Hidden Sector Dirac Dark Matter, Stueckelberg Z' Model and the CDMS
  Experiment,''
  arXiv:1003.4611 [hep-ph].



\bibitem{Fitzpatrick:2010em}
  A.~L.~Fitzpatrick, D.~Hooper and K.~M.~Zurek,
  ``Implications of CoGeNT and DAMA for Light WIMP Dark Matter,''
  Phys.\ Rev.\  D {\bf 81} (2010) 115005
  [arXiv:1003.0014 [hep-ph]].

  
  \bibitem{Tytgat}
   S.~Andreas, C.~Arina, T.~Hambye, F.~S.~Ling and M.~H.~G.~Tytgat,
  ``A light scalar WIMP through the Higgs portal and CoGeNT,''
  Phys.\ Rev.\  D {\bf 82} (2010) 043522
  [arXiv:1003.2595 [hep-ph]];
  S.~Andreas, T.~Hambye and M.~H.~G.~Tytgat,
  ``WIMP dark matter, Higgs exchange and DAMA,''
  JCAP {\bf 0810} (2008) 034
  [arXiv:0808.0255 [hep-ph]];
 M.~H.~G.~Tytgat,
  arXiv:1012.0576 [hep-ph].

  
    
 \bibitem{Tytgatbis}
  C.~Arina and M.~H.~G.~Tytgat,
  ``Constraints on Light WIMP candidates from the Isotropic Diffuse Gamma-Ray
  Emission,''
  JCAP {\bf 1101} (2011) 011
  [arXiv:1007.2765 [astro-ph.CO]];
 V.~Barger, Y.~Gao, M.~McCaskey and G.~Shaughnessy,
  ``Light Higgs Boson, Light Dark Matter and Gamma Rays,''
  Phys.\ Rev.\  D {\bf 82} (2010) 095011
  [arXiv:1008.1796 [hep-ph]].
  
  
  
  
  























 \bibitem{Mambrini:2009ad}
  Y.~Mambrini,
  ``A clear Dark Matter gamma ray line generated by the Green-Schwarz
  mechanism,''
  JCAP {\bf 0912}, 005 (2009)
  [arXiv:0907.2918 [hep-ph]];

\bibitem{Dudasline}
  E.~Dudas, Y.~Mambrini, S.~Pokorski and A.~Romagnoni,
  ``(In)visible Z' and dark matter,''
  JHEP {\bf 0908}, 014 (2009)
  [arXiv:0904.1745 [hep-ph]];



\bibitem{Higgsspace}
  C.~B.~Jackson, G.~Servant, G.~Shaughnessy, T.~M.~P.~Tait and M.~Taoso,
  ``Higgs in Space!,''
  JCAP {\bf 1004}, 004 (2010)
  [arXiv:0912.0004 [hep-ph]].
  
  
  \bibitem{Vertongen:2011mu}
  G.~Vertongen and C.~Weniger,
  ``Hunting 1-500 GeV Dark Matter Gamma-Ray Lines with the Fermi LAT,''
  arXiv:1101.2610 [hep-ph].


\bibitem{Kumar:2006gm}
  J.~Kumar and J.~D.~Wells,
  ``LHC and ILC probes of hidden-sector gauge bosons,''
  Phys.\ Rev.\  D {\bf 74}, 115017 (2006)
  [arXiv:hep-ph/0606183].



\bibitem{Antoniadis:2009ze}
  I.~Antoniadis, A.~Boyarsky, S.~Espahbodi, O.~Ruchayskiy and J.~D.~Wells,
  ``Anomaly driven signatures of new invisible physics at the Large Hadron
  Collider,''
  Nucl.\ Phys.\  B {\bf 824}, 296 (2010)
  [arXiv:0901.0639 [hep-ph]].
  
  
\bibitem{Hambye:2008bq}
  T.~Hambye,
  ``Hidden vector dark matter,''
  JHEP {\bf 0901} (2009) 028
  [arXiv:0811.0172 [hep-ph]];
  T.~Hambye and M.~H.~G.~Tytgat,
  ``Confined hidden vector dark matter,''
  Phys.\ Lett.\  B {\bf 683} (2010) 39
  [arXiv:0907.1007 [hep-ph]].


\bibitem{Hambye:2010zb}
  T.~Hambye,
  ``On the stability of particle dark matter,''
  arXiv:1012.4587 [hep-ph].










\bibitem{Baumgart:2009tn}
  M.~Baumgart, C.~Cheung, J.~T.~Ruderman, L.~T.~Wang and I.~Yavin,
  ``Non-Abelian Dark Sectors and Their Collider Signatures,''
  JHEP {\bf 0904} (2009) 014
  [arXiv:0901.0283 [hep-ph]].




















  
  
  
  
  




  
  
  
  
 















\bibitem{Fayet:2007ua}
  P.~Fayet,
  ``U-boson production in e+ e- annihilations, psi and Upsilon decays, and
  light dark matter,''
  Phys.\ Rev.\  D {\bf 75} (2007) 115017
  [arXiv:hep-ph/0702176].
  
  \bibitem{Teubner:2010ah}
  T.~Teubner, K.~Hagiwara, R.~Liao, A.~D.~Martin and D.~Nomura,
  ``Update of g-2 of the muon and Delta alpha,''
  arXiv:1001.5401 [hep-ph].
  
  
  
  
  \bibitem{Nakamura:2010zzi}
  K.~Nakamura {\it et al.}  [Particle Data Group],
  ``Review of particle physics,''
  J.\ Phys.\ G {\bf 37}, 075021 (2010).
  
 

\bibitem{EWPT}
  K.~S.~Babu, C.~F.~Kolda and J.~March-Russell,
  ``Implications of a charged-current anomaly at HERA,''
  Phys.\ Lett.\ B {\bf 408}, 261 (1997)
  [hep-ph/9705414];
  B.~Batell, M.~Pospelov and A.~Ritz,
  ``Exploring Portals to a Hidden Sector Through Fixed Targets,''
  Phys.\ Rev.\  D {\bf 80} (2009) 095024
  [arXiv:0906.5614 [hep-ph]];
  B.~Batell, M.~Pospelov and A.~Ritz,
  ``Probing a Secluded U(1) at B-factories,''
  Phys.\ Rev.\  D {\bf 79} (2009) 115008
  [arXiv:0903.0363 [hep-ph]].

\bibitem{Hook}
  A.~Hook, E.~Izaguirre and J.~G.~Wacker,
  ``Model Independent Bounds on Kinetic Mixing,''
  arXiv:1006.0973 [hep-ph].
  
  

  \bibitem{Micromegas}
  G.~Belanger, F.~Boudjema, A.~Pukhov and A.~Semenov,
  ``micrOMEGAs : a tool for dark matter studies,''
  arXiv:1005.4133 [hep-ph];
 G.~Belanger, F.~Boudjema, A.~Pukhov and A.~Semenov,
 ``Dark matter direct detection rate in a generic model with micrOMEGAs2.1,''
  Comput.\ Phys.\ Commun.\  {\bf 180}, 747 (2009)
  [arXiv:0803.2360 [hep-ph]];
   G.~Belanger, F.~Boudjema, A.~Pukhov and A.~Semenov,
  ``micrOMEGAs 2.0.7: A program to calculate the relic density of dark matter
  in a generic model,''
  Comput.\ Phys.\ Commun.\  {\bf 177}, 894 (2007).




\bibitem{WMAP}
  D.~N.~Spergel {\it et al.}  [WMAP Collaboration],
  ``Wilkinson Microwave Anisotropy Probe (WMAP) three year results:
  Implications for cosmology,''
  Astrophys.\ J.\ Suppl.\  {\bf 170} (2007) 377
  [arXiv:astro-ph/0603449];
E.~Komatsu {\it et al.}  [WMAP Collaboration],
  ``Five-Year Wilkinson Microwave Anisotropy Probe (WMAP)
  Observations:Cosmological Interpretation,''
  arXiv:0803.0547 [astro-ph].

  



  
  \bibitem{X}
 J.~I.~Collar and D.~N.~McKinsey,
  ``Comments on 'First Dark Matter Results from the XENON100 Experiment',''
  arXiv:1005.0838 [astro-ph.CO];
  T.~X.~Collaboration,
  ``Reply to the Comments on the XENON100 First Dark Matter Results,''
  arXiv:1005.2615 [astro-ph.CO];
  J.~I.~Collar and D.~N.~McKinsey,
  ``Response to arXiv:1005.2615,''
  arXiv:1005.3723 [astro-ph.CO];
  C.~Savage, G.~Gelmini, P.~Gondolo and K.~Freese,
  ``XENON10/100 dark matter constraints in comparison with CoGeNT and DAMA:
  examining the Leff dependence,''
  arXiv:1006.0972 [astro-ph.CO];
 J.~I.~Collar,
  ``Comments on arXiv:1006.0972 'XENON10/100 dark matter constraints in
  comparison with CoGeNT and DAMA: examining the Leff dependence',''
  arXiv:1006.2031 [astro-ph.CO];
  
 
\bibitem{Aprile:2011hx}
  E.~Aprile {\it et al.}  [XENON100 Collaboration],
  ``Likelihood Approach to the First Dark Matter Results from XENON100,''
  arXiv:1103.0303 [hep-ex];
 E.~Aprile {\it et al.}  [XENON100 Collaboration],
  ``First Dark Matter Results from the XENON100 Experiment,''
  arXiv:1005.0380 [astro-ph.CO].





\bibitem{XENON1T}
E. Aprile, The XENON Dark Matter Search, WONDER
Work- shop, LNGS, March 22, 2010.

\bibitem{SuperCDMS}
B. Cabrera, "SuperCDMS Development Project", 2005.

\bibitem{Angle:2011th}
  J.~Angle {\it et al.},
  ``A search for light dark matter in XENON10 data,''
  arXiv:1104.3088 [astro-ph.CO].
  
  \bibitem{Aprile:2011hi}
  E.~Aprile {\it et al.}  [XENON100 Collaboration],
  arXiv:1104.2549 [astro-ph.CO].





\bibitem{Abdo:2010ex}
  A.~A.~Abdo, M.~Ackermann, M.~Ajello, W.~B.~Atwood, L.~Baldini, J.~Ballet, G.~Barbiellini, D.~Bastieri {\it et al.},
  ``Observations of Milky Way Dwarf Spheroidal galaxies with the Fermi-LAT detector and constraints on Dark Matter models,''
  Astrophys.\ J.\  {\bf 712 } (2010)  147-158.
  [arXiv:1001.4531 [astro-ph.CO]].

\bibitem{NFW}
  J.~F.~Navarro, C.~S.~Frenk and S.~D.~M.~White,
``The Structure of Cold Dark Matter Halos'',
{\it Astrophys. J.}  {\bf 462} (1996) 563
  [arXiv:astro-ph/9508025];
 J.~F.~Navarro, C.~S.~Frenk and S.~D.~M.~White,
``A Universal Density Profile from Hierarchical Clustering'',
{\it Astrophys. J.}  {\bf 490} (1997) 493
  [arXiv:astro-ph/9611107].



\bibitem{Garde:2011wr}
  M.~L.~Garde,
  ``Constraining dark matter signal from a combined analysis of Milky Way
  satellites using the Fermi-LAT,''
  arXiv:1102.5701 [astro-ph.HE].
  
  \bibitem{Bernal:2008zk}
  N.~Bernal, A.~Goudelis, Y.~Mambrini and C.~Munoz,
  ``Determining the WIMP mass using the complementarity between direct and
  indirect searches and the ILC,''
  JCAP {\bf 0901}, 046 (2009)
  [arXiv:0804.1976 [hep-ph]].

  
  \bibitem{Goudelis:2009zz}
  A.~Goudelis, Y.~Mambrini and C.~Yaguna,
  ``Antimatter signals of singlet scalar dark matter,''
  JCAP {\bf 0912}, 008 (2009)
  [arXiv:0909.2799 [hep-ph]].


\bibitem{Conrad:2011na}
  J.~Conrad,
  ``Indirect detection of Dark matter with gamma-rays - status and
  perspectives,''
  arXiv:1103.5638 [astro-ph.CO].


\bibitem{Abramowski:2011hc}
  A.~Abramowski {\it et al.}  [H.E.S.S.Collaboration],
  ``Search for a Dark Matter annihilation signal from the Galactic Center halo
  with H.E.S.S,''
  arXiv:1103.3266 [astro-ph.HE].


\bibitem{Crocker:2010gy}
  R.~M.~Crocker, N.~F.~Bell, C.~Balazs and D.~I.~Jones,
  ``Radio and gamma-ray constraints on dark matter annihilation in the Galactic
  center,''
  Phys.\ Rev.\  D {\bf 81}, 063516 (2010)
  [arXiv:1002.0229 [hep-ph]].


\bibitem{Boehm:2010kg}
  C.~Boehm, J.~Silk and T.~Ensslin,
  ``Radio observations of the Galactic Centre and the Coma cluster as a probe
  of light dark matter self-annihilations and decay,''
  arXiv:1008.5175 [astro-ph.GA];
 C.~Boehm and P.~Uwer,
  ``Revisiting bremsstrahlung emission associated with light dark matter
  annihilations,''
  arXiv:hep-ph/0606058;
 C.~Boehm, T.~A.~Ensslin and J.~Silk,
  ``Are light annihilating dark matter particles possible?,''
  J.\ Phys.\ G {\bf 30} (2004) 279
  [arXiv:astro-ph/0208458].


\bibitem{Borriello:2008gy}
  E.~Borriello, A.~Cuoco and G.~Miele,
  ``Radio constraints on dark matter annihilation in the galactic halo and its
  substructures,''
  Phys.\ Rev.\  D {\bf 79}, 023518 (2009)
  [arXiv:0809.2990 [astro-ph]].

\bibitem{Ishiwata:2008qy}
  K.~Ishiwata, S.~Matsumoto and T.~Moroi,
  ``Synchrotron Radiation from the Galactic Center in Decaying Dark Matter
  Scenario,''
  Phys.\ Rev.\  D {\bf 79}, 043527 (2009)
  [arXiv:0811.4492 [astro-ph]].




\bibitem{Zaharijas:2010ca}
  A.~A.~Abdo {\it et al.}  [Fermi-LAT Collaboration],
  ``Constraints on Cosmological Dark Matter Annihilation from the Fermi-LAT
  Isotropic Diffuse Gamma-Ray Measurement,''
  JCAP {\bf 1004}, 014 (2010)
  [arXiv:1002.4415 [astro-ph.CO]];
  G.~Zaharijas, A.~Cuoco, Z.~Yang and J.~Conrad  [for the Fermi-LAT
                  collaboration and for the Fermi-LAT collaboration and  an],
  ``Constraints on the Galactic Halo Dark Matter from Fermi-LAT Diffuse
  Measurements,''
  arXiv:1012.0588 [astro-ph.HE];
  B.~Anderson,
  ``Fermi-LAT constraints on diffuse Dark Matter annihilation from the Galactic
  Halo,''
  arXiv:1012.0863 [hep-ph].




\bibitem{Hutsi:2011vx}
  G.~Hutsi, J.~Chluba, A.~Hektor and M.~Raidal,
  arXiv:1103.2766 [astro-ph.CO].





















\end{thebibliography}
\end{document}